# A Recipe for Galaxy Formation

Shaun Cole, Alfonso Aragón-Salamanca, Carlos S. Frenk,
Julio F. Navarro and Stephen E. Zepf[†] *Department of Physics, University of Durham, Science Laboratories, South Road, Durham DH1 3LE.*

e-mail: SHAUN.COLE@DURHAM.AC.UK, AAS@STAR.DUR.AC.UK,
C.S.FRENK@DURHAM.AC.UK, J.F.NAVARRO@DURHAM.AC.UK,
ZEPF@UCBAST.BERKELEY.EDU

**Summary.** We present a detailed prescription for how galaxy formation can be modelled in hierarchical theories of structure formation. Our model incorporates the formation and merging of dark matter halos, the shock heating and radiative cooling of baryonic gas gravitationally confined in these halos, the formation of stars regulated by the energy released by evolving stars and supernovae, the merging of galaxies within dark matter halos, and the spectral evolution of the stellar populations that are formed. The procedure that we describe is very flexible and can be applied to any hierarchical clustering theory. Our prescriptions for regulated star formation and galaxy mergers are motivated and constrained by numerical simulations. We are able to predict galaxy numbers, luminosities, colours and circular velocities. This investigation is restricted to the standard cold dark matter (CDM) cosmology and we explore the effects of varying other assumptions including the stellar initial mass function, star formation rates and galaxy merging. The results of these models we compare with an extensive range of observational data, including the $B$ and $K$ galaxy luminosity functions, galaxy colours, the Tully-Fisher relation, faint galaxy number counts, and the redshift distribution at $B \approx 22$. This combination of observed galaxy properties strongly constrains the models and enables the relative importance of each of the physical processes included to be assessed. We present a broadly successful model defined by a plausible choice of parameters. This fiducial model produces a much more acceptable luminosity function than most previous studies. This is achieved through a modest rate of galaxy mergers and strong suppression of star formation in halos of low circular velocity by energy injected by supernovae and evolving stars. The model also accounts for the observed faint galaxy counts in both the $B$- and $K$-bands and their redshift distributions. However, it fails to produce galaxies as red as many observed ellipticals and, compared with the observed Tully-Fisher relation, the model galaxies have circular velocities which are too large for their luminosities.

## 1 Introduction

Theoretical studies of galaxy formation require an understanding of many diverse astrophysical processes. For example, in hierarchical clustering models, galaxy formation is driven by the dynamics of an evolving population of dark matter halos within which gas cools and turns into stars. Processes associated with star formation and evolution may, in turn, feed back into the large-scale behaviour of the mass and radiation fields. If they are to have any prospect of being realistic, models of galaxy formation must therefore include, at least, a treatment of the following physical effects:

*(i)* The evolution of the dark matter distribution in its proper cosmological setting, including the non-linear processes associated with the collapse and merging of dark matter halos.

*(ii)* The dynamical behaviour of gas coupled gravitationally to the dark matter and subject to shocks, cooling and heating processes.

*(iii)* The evolution of stellar populations, including their spectrophotometric properties as a function of time and the injection of energy, mass, and metals into the surrounding interstellar and intergalactic gas.

*(iv)* Tidal interactions and mergers of individual galaxies in a dynamically active environment.

---

[†] Present Address: Department of Astronomy, University of California, Berkeley CA. 94720, U.S.A.

Of all these processes, the least uncertain are the evolution of dark matter halos and the evolution of galactic stellar populations. The former has been investigated in considerable detail using N-body simulations (For a review see Frenk 1991 and references therein) and analytic techniques (Press & Schechter 1974; Bardeen et al. 1986 ; Bond et al. 1991; Bower 1991, Lacey & Cole 1993, Kauffmann & White 1993). The spectrophotometric evolution of a stellar population formed with a given initial mass function can now be predicted with some reliability, at least for the case of solar metallicity (e.g. Bruzual & Charlot 1993).

By far the least understood processes are those associated with gas dynamics and star formation. The importance of radiative cooling was first recognized by Rees & Ostriker (1977), Binney (1977), and Silk (1977). Cooling is very efficient on galactic and sub-galactic scales and, in the absence of heating sources, the gas is expected to quickly lose its pressure support and collapse to the centre of a virialized halo where it will eventually turn into stars. The products of stellar formation and evolution influence the cooling properties of the gas and thus regulate the supply of cold gas and the subsequent efficiency of star formation (White & Rees 1978; Cole 1991; White & Frenk 1991). N-body/hydrodynamic simulations are now beginning to address some of these issues (Katz & Gunn 1991; Navarro & Benz 1991; Navarro & White 1993; Katz et al. 1992; Evrard et al. 1993; Katz & White 1993), as well as the related question of the different dynamical histories of dark halos and the galaxies they harbour (Cen & Ostriker 1992, Katz & White 1993, Evrard et al. 1993, Navarro, Frenk & White 1994). There is now quantitative evidence that the merging timescale of a dissipative component can substantially exceed that of its collisionless halo, but reliable calculations of the merger rates of galaxies in different environments are still some way away.

In spite of the many remaining uncertainties, the body of knowledge accumulated so far is sufficiently extensive to justify building detailed models of galaxy formation incorporating as many of the salient physical processes as possible. Modern approaches to this problem have their roots in the pioneering work of White & Rees (1978) and Larson (1974). The first attempts to build upon these ideas taking advantage of the experience gathered in the 1980s, particularly in connection with cosmological simulation work, were those of Cole (1991), White and Frenk 1991 (hereafter WF), and Lacey & Silk (1991). These papers laid out analytic frameworks which can readily accommodate different assumptions concerning the initial spectrum of mass fluctuations, the dynamics of prestellar and processed gas, the effects of stellar evolution, etc. They allow detailed predictions to be made for the time evolution of the galaxy population and its properties: star formation rates; luminosity functions; relations between circular velocity, luminosity, metallicity and mass-to-light ratios; counts of faint blue galaxies as a function of luminosity and redshift, etc. More recently, Lacey et al. (1993) have constructed models which, while similar in spirit to those of Cole and WF, differ in a number of important respects such as the inclusion of a population synthesis model, the treatment of the dark matter distribution (for which they employ the peak, rather than the Press-Schechter formalism) and the star formation law (which they relate to tidal interactions rather than to the cooling timescale within individual halos).

In spite of their many differences of detail, a number of general conclusions have emerged from these "first generation" models. For example, as emphasized by Cole (1991) and WF, efficient heating of primordial gas is required in cosmogonies in which structure grows through hierarchical clustering (not just CDM) to prevent most of the gas from condensing early on into halos at the lowest level of the hierarchy. The galaxy luminosity function in some of the models, particularly those of Lacey et al. (1993), approximates a Schechter function –a notable success – but its faint end is much steeper than that of the observed field luminosity function (e.g. Loveday et al. 1992). The WF and Lacey et al. models reproduce quite well the properties of the faint blue counts, but the extent to which this success is tied to the failure to reproduce the faint luminosity function is unclear. The WF models also reproduce the characteristic luminosities of galaxy clusters and give rise to a bias in the distribution of galaxies which has the correct sign but insufficient amplitude to reconcile the observed kinematics of galaxy clustering with the CDM assumption of a flat universe. The circular velocities and luminosities of the galaxies in the WF and Cole models obey a relationship which has a form similar to the observed Tully-Fisher relation (Tully & Fisher 1977), and relatively small scatter, but the circular velocities at a given luminosity are typically about a factor of two too large. This problem is also present in a different guise in the Lacey et al. (1993) study. Finally, as these last authors emphasize, models in which structure builds up hierarchically have a built-in difficulty in making luminous galaxies (which typically form late) with the red colours observed for the brightest ellipticals.

An important extension of this approach was recently carried out by Kauffmann et al. (1993). In an elegant paper, they grafted many of the techniques developed by WF into a Monte-Carlo implementation of the Press-Schechter formalism which allowed them to follow the merging histories of individual structures as they form in the hierarchical clustering process. By including a population synthesis model they were able to investigate the detailed spectrophotometric properties of the stellar components that form. Kauffmann et al. focussed primarily on the processes that establish the different morphological types and explored environmental effects in considerable detail. They found that models of this type provide a natural explanation for the existence of the Hubble sequence and for a number of observed trends involving the relative abundances, luminosities, colours, stellar ages, gas content, and environment of galaxies of different types.

In this paper we develop new models based also on Monte-Carlo realizations of the hierarchical clustering process and the use of stellar population synthesis techniques. The way in which we implement these features, however, is very different from that of Kauffmann *et al.* Our Monte-Carlo techniques are based on the "block model" of Cole & Kaiser (1988), rather than on a direct extension of the Press-Schechter formalism, and our population synthesis models also differ. Furthermore, we treat the key processes of star formation, energy feedback, and galaxy mergers in an entirely different way from that of Kauffmann *et al.* Unlike them, we focus on the properties of the galaxy population as a whole and do not consider in any detail the distinction between different morphological types. With such different operational implementations of what is essentially the same physics, it is highly instructive to compare results.

Galaxy formation studies of the type discussed here have generally adopted the standard cold dark matter cosmogonic model (CDM; Blumenthal *et al.* 1984, Davis *et al.* 1985) to describe the abundance and the internal structure of galactic halos as a function of time. We follow this practice and concentrate on how galaxy formation is affected by changes in the parameters used to model gas dynamics and star formation. Our model therefore assumes $\Omega = 1$, $H_0 = 50 \, \text{km}\,\text{s}^{-1}\text{Mpc}^{-1}$ and the CDM fluctuation spectrum, normalized so that the amplitude of mass fluctuations within spheres of radius $8h^{-1}$Mpc is $\sigma_8 = 0.67$, corresponding to a bias parameter, $b \equiv 1/\sigma_8 = 1.5$. This value is just outside the range allowed by the masses and abundances of rich galaxy clusters (White *et al.* 1993a), but is marginally consistent with the amplitude of the microwave background temperature anisotropies measured by the COBE satellite (Smoot *et al.* 1992). If one were to add extra large scale power to this model to obtain better agreement with the COBE data and with observations of galaxy clustering on large scales (*e.g.* Maddox *et al.* 1990a), our results would be little changed. However if the power spectrum at shorter wavelengths were altered, perhaps by tilting the whole spectrum, by assuming a mixture of hot and cold dark matter, or simply by varying $\sigma_8$, this would give rise to significant changes. We plan to study such effects in a later paper.

The following section sets the background and describes the techniques we employ to model gravitational evolution (§2.1), star formation and feedback (§2.2), galaxy mergers (§2.3), and the evolution of stellar populations (§2.4) . Section 3 describes how these ingredients are meshed together into a galaxy formation algorithm. The predictions of a "fiducial model" are compared with a comprehensive range of observational data in §4.1. The effect of varying each of our model parameters is studied in §4.2, while §4.3 presents some testable predictions for which relevant data do not yet exist. We discuss and summarize our results in §5. This is a long paper and some readers may wish to skip technical details on a first reading. For those who wish to do so, we suggest skipping through section 2, pausing only at Figure 2 and equations (2.5) to (2.11) which define our prescription for star formation and its regulation by the energy liberated by supernovae and evolving stars.

## 2 Techniques

Our model of galaxy formation can be divided into four distinct elements. First (§2.1), we present a model for the hierarchical evolution via accretion and mergers of the population of dark matter (DM) halos and a simple model of the structure of each DM halo. Second (§2.2), we model the physical processes which act on the baryonic component of each halo: shock heating, radiative cooling, and star formation. Third (§2.3), we model the merger rate between galaxies within DM halos. Last (§2.4), we use the techniques of stellar population synthesis to follow the evolution of the stars that form and derive the observable properties of our model galaxies.

### 2.1 GRAVITATIONAL EVOLUTION

The structure in the universe today is usually assumed to have grown from small-amplitude density fluctuations generated by physical processes in the early universe. In hierarchical clustering theories (such as CDM), primordial fluctuations are amplified by gravitational instability until they become nonlinear, giving rise to bound clumps which become progressively more massive as they merge together and accrete surrounding material. A simple analytic calculation of the abundance of clumps and their history of hierarchical merging has been performed by Cole (1989), Bond *et al.* (1991), Bower (1991), and Lacey & Cole (1993) and appears to be in surprisingly good quantitative agreement with the hierarchical mergers synthesized in cosmological N-body simulations (Efstathiou *et al.* 1988; Frenk *et al.* 1988). The evolving mass function of halos turns out to be identical to that derived by Press & Schechter (1974), and so in one sense this description can be viewed as an extension of the Press-Schechter theory which enables halo merger rates to be quantified explicitly. Here we will use a Monte-Carlo approach to construct a series of realizations whose ensemble properties approximately reproduce the mass function and merger history described by the analytic theory. This is based on the "block model" introduced by Cole & Kaiser (1988) to study the evolution of galaxy groups and clusters (see also Cole 1991).

The basic input from a cosmological model is the power spectrum of density fluctuations which is used to compute the variance of the density fluctuations, $\sigma(M)$, as a function of mass scale. We take a large block of the universe containing mass $M$, typically $10^{16} M_\odot$, and assign to this volume a density perturbation, $\delta$, drawn from a gaussian distribution with zero mean and variance $\sigma(M)$. We then divide this block in two producing two daughter blocks. To one of these daughters we add an extra gaussian perturbation to that which it inherits from its parent. The variance of the gaussian from which this perturbation is drawn is such that when added in quadrature to $\sigma(M)$ it produces the variance $\sigma(M/2)$. We subtract this same perturbation from the other daughter. This produces two volumes with gaussian distributed density perturbations with variance $\sigma(M/2)$ and whose mean density perturbation equals that of the parent block that they make up. We apply this procedure repeatedly to each daughter until we have divided the original block into $2^N$ volumes each of mass $M/2^N$. These $2^N$ volumes have a distribution of density perturbations, $\delta$, which is gaussian with variance $\sigma(M/2^N)$ and whose individual densities are consistent with their $2^{N-1}$ parents one level up in the hierarchy of mass resolution.

We typically construct a hierarchy of 20 levels making the mass resolution of the lowest level $10^{16}/2^{19} = 1.9 \times 10^{10} M_\odot$. In what follows we explicitly check that our model predictions are insensitive to this resolution limit. This hierarchy is then converted into a merger history of all the material contained within the block. For each volume in each level of the hierarchy the redshift, $z$, at which the linear theory perturbation grows to a critical value, $\delta_c$, may be calculated. For a critical density universe ($\Omega = 1$), the linear growth rate is simply proportional to the expansion factor, $\delta(t) \propto a = (1+z)^{-1}$ and the threshold for collapse of a spherically symmetric overdense region is $\delta_c = 1.686$ (*e.g.* Gunn & Gott 1972). This redshift will be the redshift at which this volume collapses to form a virialized halo, *unless* one of its parents higher up in the hierarchy collapses at an earlier redshift. This procedure closely mimics the behaviour of virialized halos in N-body simulations, where substructure is rapidly erased after collapse (Frenk *et al.* 1988). By tracing the sequence of collapse redshifts throughout the block we can follow the sequence of mergers starting from the formation of halos of the lowest resolved mass up to the formation of the present generation of halos. Thus, this Monte-Carlo model gives us the formation redshift of each halo and its eventual fate in terms of which halos it merges with and at what later redshift.

The analytic expression for the number density of halos in the mass interval $M$ to $M + dM$ is

$$\frac{dn}{dM}(M,z)\, dM = \left(\frac{2}{\pi}\right)^{1/2} \frac{\rho_0}{M^2} \frac{\delta_c(1+z)}{\sigma(M)} \left|\frac{d\ln\sigma}{d\ln M}\right| \exp\left[-\frac{\delta_c^2(1+z)^2}{2\sigma^2(M)}\right] dM, \qquad (2.1)$$

where $\rho_0$ is the mean mass density of the universe (Press & Schechter 1974). Figure 3 of Cole & Kaiser (1988) compares this mass function with that arising in the block model for the case of the CDM power spectrum. The mass functions have an exponential cutoff above a characteristic mass, $M^\star(z)$, defined by $\sigma(M^\star) = \delta_c(1+z)$, which increases with decreasing redshift. At masses which are low compared to $M^\star$, the mass functions have a power law form. This and related analytic formulae have been found to match the results of cosmological N-body simulations remarkably well (Efstathiou *et al.* 1988; Bond *et al.* 1991; Lacey & Cole, in preparation). At all epochs, over the mass range reliably modelled, the number of halos identified in the simulations differs by less than 25% from the number given by equation (2.1). Nevertheless, it should be noted that the dynamic range of these simulations is limited and therefore little is known about the validity or otherwise of the power law form of the mass function for masses $M \lesssim M^\star/10$.

The ability to follow the individual histories of the halos is the main advantage of the Monte-Carlo method over the statistical approach employed by WF. The block model provides an adequate description of the evolution of the halo population, but provides no information about the internal structure of any of the halos. To first approximation, the DM density profiles of virialized halos found in CDM cosmological N-body simulations are well described by an "isothermal" density profile, $\rho_{\rm DM} \propto r^{-2}$ (Quinn *et al.* 1986; Frenk *et al.* 1988). The flat rotation curve implied by this density profile is consistent with the observed HI rotation curves of spiral galaxies. Application of the virial theorem to the collapse of a uniformly overdense spherical perturbation predicts that the mean density contrast of the virialized region of a halo is approximately 200 times the background density at the time of virialization (*e.g.* Lacey & Cole 1993). The radius defined by this density contrast is also found to delineate the transition from random virial velocities in the interior to ordered inflow from the surroundings. The mass of the halo (inside this radius) is then simply related to the halo circular velocity and formation redshift by

$$M_{\rm halo} = 2.35 \times 10^5 \left(\frac{V_c}{\rm km\,s^{-1}}\right)^3 (1+z)^{-3/2} \,{\rm M}_\odot/h. \tag{2.2}$$

2.2 THE INTERSTELLAR MEDIUM AND STAR FORMATION

In contrast with the significant advances in our understanding of the evolution of the gravitationally dominant component of the universe, the fate of the baryonic component still remains an unresolved issue. However, it is clear that a successful description of the evolution of gas on galactic scales needs to include a number of key ingredients. The most important are the ability of gas to radiate away a large fraction of its energy (Rees & Ostriker 1977; Binney 1977; Silk 1977), together with the formation of stars and the subsequent effect that evolving stars and exploding supernovae have on their surroundings. In this section, we use both theoretical considerations and the results of numerical simulations to derive simple scaling laws which relate these processes to the potential well of a galactic halo.

2.2.1 Radiative Cooling

The fraction of the baryonic content of a halo which can actually cool and reach the centre is determined by the balance between the cooling and dynamical timescales of the system, which in turn is fixed by the density and temperature structure of the gas component. In the absence of radiative cooling, numerical simulations show that the gas closely follows the spatial distribution of the dark matter in a virialized system (Evrard 1990). Shocks are very effective at transforming the gas kinetic energy into heat, and the gas temperature is rapidly raised to the virial temperature of the halo,

$$kT_{\rm vir} = \frac{1}{2}\mu m_p V_c^2. \tag{2.3}$$

Here $\mu$ is the mean molecular weight, and $m_p$ and $k$ are the proton mass and Boltzmann's constant, respectively. Together with the assumption that the dark matter density profile can be approximated by an isothermal sphere, this fully specifies the density-temperature distribution of the diffuse gas component in each halo, $\rho_{\rm gas}(r) \propto \rho_{\rm DM}(r) \propto r^{-2}$, and $T_{\rm gas} = T_{\rm vir}$.

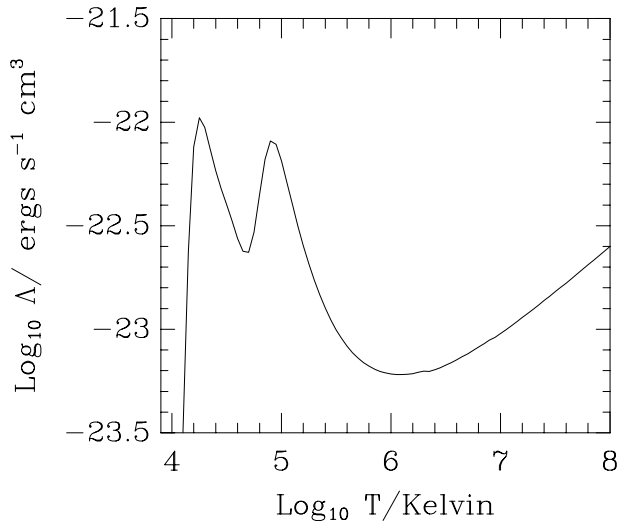

**Figure 1.** The cooling function, $\Lambda(T)$, for a primordial mixture of 77% Hydrogen and 23% Helium in collisional ionisation equilibrium computed by Sutherland & Dopita (1993).

Once the gas density and temperature structure have been fixed, we can compute the amount of mass that can cool in a halo of circular velocity, $V_c$. At each radius, we define the cooling timescale as the ratio of the specific thermal energy content of the gas and the cooling rate per unit volume, $n_e^2 \Lambda(T)$:

$$\tau_{\rm cool}(r) = \frac{3}{2} \frac{\rho_{\rm gas}(r)}{\mu m_p} \frac{kT_{\rm gas}}{n_e^2(r)\Lambda(T_{\rm gas})}, \qquad (2.4)$$

where $n_e(r)$ is the electron number density. We adopt the cooling function given by Sutherland & Dopita (1993) for a gas with primordial abundances, as plotted in Figure 1. We assume that the amount of gas that can cool is the mass initially enclosed by the "cooling radius" $r_{\rm cool}$ – the radius where $\tau_{\rm cool}$ equals the *lifetime* of the system, $t_l$. The Monte Carlo model described in §2.1 provides a simple definition of this lifetime as the time between the formation of the halo and its subsequent incorporation into a new halo of twice or more its original mass. At this point we assume that any remaining hot gas will be shock heated to the virial temperature of the new halo. Since cooling is quite effective at high redshift, it may happen that $r_{\rm cool}$ is actually larger than the virialized region of the halo under consideration. In this case, we take the cooled mass to be the total gas mass of the halo.

### 2.2.2 Star Formation and Feedback

The cold dense gas that accumulates at the centres of halos will eventually start forming stars. The baryonic content of a halo is therefore distributed among three separate components: a hot diffuse atmosphere; cold, dense, star-forming condensations; and stars. In the previous section we saw how radiative cooling causes gas to be transferred from the hot to the cold component. We now consider how star formation and evolution couple all three components. In order to determine how to distribute the baryons between the three "phases" we construct a set of equations relating the star formation rate to the total amount of cooled gas and the fraction that is ejected from the cold phase back into the hot diffuse atmosphere. We write these equations as simple scaling laws depending solely on the circular velocity of the halo being considered.

The particular choice for these scaling laws is motivated by the numerical simulations of Navarro & White (1993), who used a 3D Lagrangian hydrodynamical code to model the formation of gaseous disks following the collapse of a rotating sphere containing a mixture of gas and dark matter. In an attempt to mimic the effect of supernova (SN) explosions on the interstellar medium, the formation of stars in their simulations is accompanied by the injection of energy into the surrounding gas particles. This energy is assumed to affect only the temperature and velocity field of the gas. A complete account of the simulations and details of the numerical procedure are given in Navarro & White (1993).

Navarro and White adopt an IMF that results in $4 \times 10^{48}$ ergs being released by supernovae for every $1 M_\odot$ of gas turned into stars and define $f_v$ to be the fraction of this energy that is dumped as kinetic energy. The simulations show that $f_v$ is the single most important parameter governing the evolution of a gaseous disk actively forming stars. Setting $f_v = 0$ implies that SN can only raise the local temperature of the surrounding gas. Since stars form in high density regions, where cooling is extremely efficient, SN have practically no influence on the surrounding gas when $f_v = 0$. On the other hand, values of $f_v$ as modest as 0.1 have a dramatic effect on the subsequent evolution of the gas, some of which is devolved to the hot diffuse component by "winds" driven by supernovae and young evolving stars (Bressan *et al.* 1993) . For a given value of $f_v$, both the mean star formation rate and the amount of gas pushed out of the disk are tightly related to the depth of the potential well of the halo. In this paper we consider models with a variety of different IMFs and therefore $f_v$ should not be viewed strictly as a fractional energy, but simply as the amount of energy from supernovae and stellar winds dumped into the interstellar medium as kinetic motion in units of $4 \times 10^{48}$ ergs per $1\,M_\odot$ of stars formed. For brevity, we shall refer to the energy released by all processes associated with stellar evolution as "supernova energy" or "feedback".

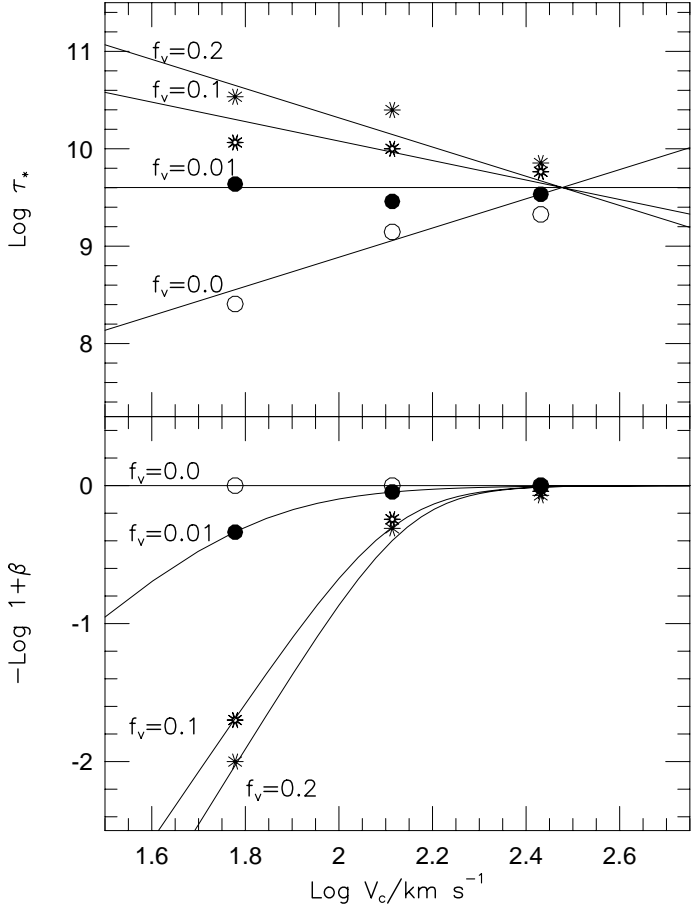

**Figure 2.** The efficiency of star formation as a function of the circular velocity of the surrounding dark matter halo in the numerical simulations of Navarro & White 1993. a) The logarithm of star formation timescale, $\tau_\star$, in years, for four different values of the feedback parameter $f_v$. The symbols show the peak star formation rates from the simulations and the solid lines show the simple power law models (equation (2.10)) we use to characterize these results. The resulting parameter values, $\alpha_\star$, of these models are given in Table 1. b) The logarithm of the fraction of gas that remains in the cold phase and is available to form stars, $\log_{10}(1-f_{\rm hot}) = -\log_{10}(1+\beta)$ (equation (2.11)), as a function of circular velocity for four different values of the feedback parameter $f_v$. The curves show the fit of equation (2.11) with parameter values $V_{\rm hot}$ and $\alpha_{\rm hot}$ given in Table 1.

We now set out the equations which govern the star formation rates in each galaxy of our Monte Carlo simulations. The supply of cold gas capable of forming stars in a newly formed halo is regulated by the mass of hot gas that can cool over the lifetime $t_l$ of a halo, as described in the previous subsection, and by the merger of the cold gas reservoirs of two or more existing galaxies. (Galaxy mergers are discussed in §2.3.) Let us call $m_c^0$ the total mass of cold gas computed this way. Note that $m_c^0$ is *not* a simple function of the halo circular velocity, as it depends on the past star formation and merger histories of the galaxy in question. It is therefore influenced by the galaxy's environment. We assume that the star formation rate is at time $t$ measured from the onset of a star formation episode given by

$$\dot{m}_\star(t, V_c) = m_c(t, V_c)/\tau_\star(V_c) = (m_c^0 - m_\star(t, V_c) - m_{\rm hot}(t, V_c))/\tau_\star(V_c), \tag{2.5}$$

where $m_{\rm hot}(t, V_c)$ is the mass of cooled gas reheated by the energy released from SN that is returned to the hot phase and $m_\star(t, V_c)$ is the mass of stars formed during the current star formation episode. For simplicity, we neglect the gas returned to the interstellar medium by evolving stars and supernovae. It seems appropriate to assume that the mass returned to the hot phase should be proportional to the mass of stars formed,

$$\dot m_{\rm hot}(t, V_c) = \beta(V_c)\, \dot m_\star(t, V_c). \tag{2.6}$$

Both $\tau_\star$ and $\beta$ are simple functions of $V_c$. Therefore,

$$m_\star(t, V_c) = \frac{m_c^0}{1+\beta}\left[\, 1 - \exp(-(1+\beta)\, t/\tau_\star)\,\right], \tag{2.7}$$

and

$$m_{\rm hot}(t, V_c) = \beta(V_c)\, m_\star(t, V_c), \tag{2.8}$$

so that at times long compared with $\tau_\star/(1+\beta)$, the fraction of cooled gas that is reheated is

$$f_{\rm hot}(V_c) = \frac{\beta(V_c)}{1 + \beta(V_c)}. \tag{2.9}$$

We write the star formation timescale $\tau_\star(V_c)$ and the proportionality factor $\beta(V_c)$ as simple scaling laws;

$$\tau_\star(V_c) = \tau_\star^0 \left(\frac{V_c}{300\,{\rm km\,s^{-1}}}\right)^{\alpha_\star} \tag{2.10}$$

$$\beta(V_c) = (V_c/V_{\rm hot})^{-\alpha_{\rm hot}}. \tag{2.11}$$

The four parameters $\alpha_{\rm hot}$, $V_{\rm hot}$, $\alpha_\star$, and $\tau_\star^0$ fully specify the star formation history of every galaxy in our model during its lifetime. The simulations of Navarro & White (1993) suggest that the values of all these parameters depend only on the choice of the feedback parameter $f_v$ (see their Table 2). This is shown in figure 2, where we plot $1+\beta$ and $\tau_\star$ as a function of $V_c$, after identifying the "peak" star formation rates in the simulations with the maximum star formation rates derived from (2.7).

**Table 1**

| $f_v$ | $\alpha_\star$ | $V_{\rm hot}/\,{\rm km\,s^{-1}}$ | $\alpha_{\rm hot}$ |
|---|---|---|---|
| 0.0 | 1.5 | 0.0 | – |
| 0.01 | 0.0 | 63.0 | 3.0 |
| 0.1 | -1.0 | 130.0 | 5.0 |
| 0.2 | -1.5 | 140.0 | 5.5 |

A number of important points should be noted in this figure. First, when $f_v > 0$ large amounts of cooled gas are returned to the hot diffuse component in low mass systems, where star formation is severely slowed down. This is a direct result of the low binding energy of these systems (cf Dekel and Silk 1986). Second, if $f_v \geq 0.1$, the star formation timescales can be comparable or larger than a Hubble time in low-mass systems. Third, a note of caution. The agreement between our parameterization and the results of the numerical simulations is encouraging, but not perfect –the star formation rates in the simulations are not exponentially decaying as assumed in our model. As a result, the values of $\tau_\star^0$ and $\alpha_\star$ actually depend on the details of the procedure used to fit eq. (2.7) to the results of the numerical experiments. In view of this, and in order to keep our model simple, we decided to treat $\tau_\star^0$ as a free parameter. The value of $\tau_\star^0$ required to match the observed colours of galaxies turns out to be only a factor of two different from that suggested by the power law fits shown in figure 2. Finally, let us stress again that the four quantities $\alpha_\star$, $\tau_\star^0$, $\alpha_{\rm hot}$, and $V_{\rm hot}$, are not independent of one another but rather reflect the choice of *one* single physical parameter: the efficiency with which the energy released by supernovae and evolving stars can perturb the velocity field of the interstellar medium. Note that as a result of the large values of $\alpha_{\rm hot}$ required when $f_v \geq 0.1$, the importance of feedback in our model declines more rapidly with halo circular velocity than in the prescription adopted by White & Rees (1978), WF and Kauffmann *et al.* (1993).

2.3   MERGERS AND THE FATE OF GALAXIES WITHIN DARK HALOS

Our model must also explicitly incorporate the fate of galaxies and their gas during a merger of their surrounding halos. It has long been clear that in a hierarchically clustering cosmogony mergers between galaxies must be much less frequent than those of their parent halos. A galaxy cluster is a prime example of this, where individual galaxies are thought to orbit in the overall potential of the cluster halo presumably stripped of their own individual halos. When two halos merge, their hot diffuse gaseous atmospheres

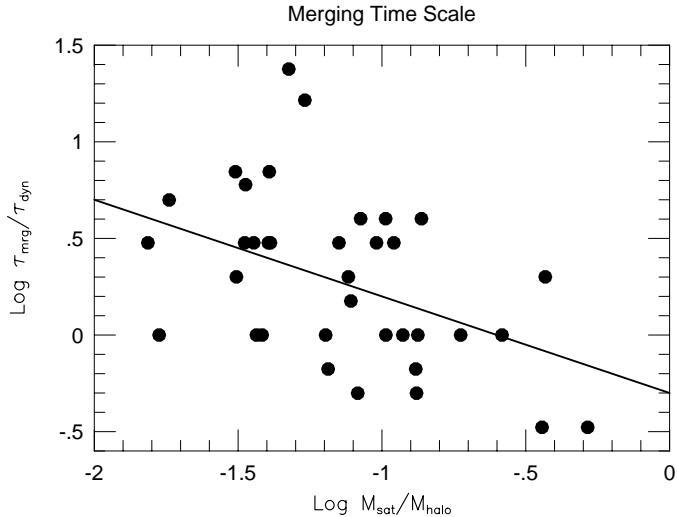

**Figure 3.** A comparison of the merger timescale given by equation (2.12), with $\alpha_{\rm mrg} = 0.25$ and $\tau^0_{\rm mrg}/\tau_{\rm dyn}=0.5$, with the results of numerical simulations (Navarro, Frenk & White 1993b)

collide, shocks convert the kinetic energy of the collision into thermal energy and produce an atmosphere in quasi-static equilibrium within the new halo potential (Evrard 1990; Navarro & White 1993; Pearce 1993). Numerical simulations show that the fate of tightly bound cores, on the other hand, is quite different from that of the more loosely bound material permeating the halo (Katz & Gunn 1991; Navarro & Benz 1991). Dense cores of gas and stars can survive the initial merger practically intact and only later collide with one another and perhaps coalesce.

There are various factors that affect the probability that a collision will occur and that the result of the collision will be a merger. Consider a halo of mass, $M_{\rm halo}$, which has just formed as the result of the merger of two or more halos. Here we will refer to the galaxies that have fallen into this halo as "satellite galaxies", although "group members" may be a more appropriate description if no one galaxy is dominant in the new halo. The first physical process acting on these satellites is dynamical friction. A satellite galaxy orbiting in the halo experiences a drag force that causes its orbit to decay and the galaxy to spiral to the centre of the halo, greatly enhancing the probability that it will undergo a collision. A simple calculation shows that the more massive the satellite, the shorter the timescale in which its orbit decays. Second, when a collision occurs between two galaxies, the probability that it will result in a merger is an increasing function of the ratio of the internal velocity dispersion of the galaxies involved to the encounter velocity (see White 1976; Aarseth & Fall 1980 and references therein). The internal velocity of the satellite galaxy is related to the total mass of the halo in which the galaxy first formed, $M_{\rm sat}$, by equation (2.2), while the orbital velocity scales as the circular velocity of the recipient halo. Hence, in a collision the typical ratio of the galaxy internal velocity dispersion to the encounter velocity is an increasing function of $M_{\rm sat}/M_{\rm halo}$. Furthermore, since dynamical friction has a larger effect on more massive satellites, we expect the probability of a satellite galaxy undergoing a merger to be an increasing function of $M_{\rm sat}/M_{\rm halo}$.

These considerations suggest that the timescale governing the merging of satellite galaxies, or group members, orbiting in a halo of mass, $M_{\rm halo}$, should be a decreasing function of $M_{\rm sat}/M_{\rm halo}$. However, a more detailed description of the probable fate of a satellite galaxy orbiting in a DM halo of a larger galaxy or group is not yet available, especially because the survival time of a satellite galaxy depends sensitively on the uncertain angular momentum of its initial orbit. We therefore define a merger timescale, $\tau_{\rm mrg}$, by the simple scaling law,

$$\tau_{\rm mrg} = \tau^0_{\rm mrg}\,(M_{\rm halo}/M_{\rm sat})^{\alpha_{\rm mrg}}, \qquad (2.12)$$

with $\alpha_{\rm mrg} > 0$. Here, $M_{\rm sat}$ is the initial mass of the halo of the infalling satellite galaxy under consideration and $M_{\rm halo}$ is the total mass of the new composite halo in which it is now orbiting. The free parameters are $\tau^0_{\rm mrg}$, which we parameterize as a fixed fraction of the dynamical time of the halo, $\tau_{\rm dyn}$, and the constant $\alpha_{\rm mrg}$, which controls the efficiency of the merger process as a function of the mass ratio. We define the dynamical time, $\tau_{\rm dyn}$, as one half of the age of the universe at the time when the halo forms. To determine appropriate values for these two free parameter we seek guidance in the results of numerical simulations which incorporate the effects discussed above.

Figure 3 compares equation (2.12) with $\alpha_{\rm mrg} = 0.25$ and $\tau^0_{\rm mrg}/\tau_{\rm dyn}=0.5$ with the results of simulations of merging satellites performed by Navarro, Frenk & White (1994). In these simulations cold dense baryonic condensations were identified during the collapse and formation of a galaxy halo. The evolution of these satellites was then followed as they orbit in and eventually merge at the centre of the galaxy halo. The symbols in Figure 3 show the time at which these mergers take place, measured from the formation time of the halo in units of the dynamical time, as a function of the satellite mass. There is a wide distribution of merger times, reflecting the variety of initial satellite orbits, but there is also a clear trend towards shorter merger times for more massive satellites in accordance with our simple model. Our scaling law is therefore an approximate model for the merging of galaxies within DM halos in the mass range considered. In order to decide whether a satellite galaxy has merged with the central galaxy of the new halo, we compare its merger timescale, $\tau_{\rm mrg}$, to the lifetime, $t_l$, of the new halo, *i.e.* the time between its formation epoch and either the required output time or the time at which the halo merges again, whichever is earliest. Galaxies for which $\tau_{\rm mrg} > t_l$ retain their own identity as satellite galaxies or cluster members. Those for which $\tau_{\rm mrg} < t_l$, merge together at the centre of the common halo and are assigned the circular velocity of this new system. In the event that no galaxies merge according to this rule, we select the one with highest circular velocity to be the dominant galaxy onto which any gas cooling during the lifetime of the new common halo will accrete.

## 2.4 STELLAR POPULATION SYNTHESIS

Given a star formation history, a stellar population synthesis model gives the observable properties of the emerging stellar population, *e.g.* its spectral energy distribution (SED), absolute magnitude and colours, as a function of time. This technique, pioneered by Tinsley (1972, 1980), is based on theoretical and observational studies of stellar evolution. The models assume that stars are born with masses distributed according to a universal initial mass function (IMF), at a rate governed by a star formation rate (SFR). The model stars are then evolved according to theoretical evolutionary tracks. Using empirical or theoretical calibrations, the predicted colours or spectra of the composite population are obtained.

Several problems arise when implementing this technique. First, poor theoretical understanding of the physics of convection and the late stages of stellar evolution after the onset of helium core burning introduce uncertainties in the treatment of horizontal branch, asymptotic giant branch (AGB) and post-AGB stars (see *e.g.* Renzini & Fusi-Peci 1988). In some stellar population synthesis models, these stellar classes are completely ignored or introduced in a semi-empirical way. A fully satisfactory procedure for dealing with these late stages of stellar evolution has not yet been developed (see, *e.g.*, Renzini 1989). Therefore, large uncertainties still exist in the spectral evolution in the near infrared, where AGB stars could play an important role, and in the ultraviolet, where the post-AGB stars might make a substantial contribution. Another major problem is the way these models deal with stars of different chemical compositions and the chemical evolution of the system. Composite stellar systems might contain stars with a wide range of metallicities. In particular, the cores of giant elliptical galaxies are believed to contain stars that are considerably more metal rich than those in the solar neighbourhood, and stellar libraries do not cover the complete range of metal abundances. In addition, theoretical stellar evolutionary tracks for metal rich stars are difficult to compute precisely because of uncertainties in the opacities, and so are less reliable than solar abundance ones.

Several authors have used population synthesis techniques with varying degrees of success, making different sets of assumptions when dealing with the problems mentioned above (see, *e.g.* , Tinsley 1972, 1980; Tinsley & Gunn 1976; Bruzual 1983; Renzini & Buzzoni 1986; Arimoto & Yoshii 1986, 1987; Guiderdoni & Rocca-Volmerange 1987, 1990; Aragón *et al.* 1987; Buzzoni 1989; Charlot & Bruzual 1991; Bruzual & Charlot 1993). Out of this wide selection, we have chosen the model of Bruzual & Charlot (1993)[†]. It includes up-to-date stellar evolution calculations that minimize inconsistencies in the physical modelling of different stellar mass ranges. The stellar evolutionary tracks cover all the relevant stages of stellar evolution, from the main sequence to the remnant stage, including AGB and post-AGB stars, although these are included in a somewhat semi-empirical way. A very important consideration is that the method employed to build the models —called by these authors "isochrone synthesis"— allows one to accurately follow the evolution of arbitrarily short bursts of star formation. An important drawback of this model, also present in most of its competitors, is the assumption of solar metallicity and hence the neglect of the effects of chemical evolution. Unfortunately, models which include chemical evolution, like those of Arimoto & Yoshii (1986, 1987), do not contain important stages of stellar evolution and are unreliable for very short bursts of star formation.

Bruzual & Charlot's models produce an acceptable fit to the 2000Å − 2$\mu$m integrated SEDs of present day galaxies of all spectral types, despite the fact that chemical evolution is ignored and solar metallicity is used. The explanation for this could be that even in giant ellipticals the average global metallicity is not far from the solar value inspite of the high metallicity of the central regions.

---

[†] The Bruzual & Charlot (1993) models include a revision of the incorrect main-sequence lifetimes in the input stellar tracks of Maeder & Meynet (1989) and correct a normalization error in the model of Charlot & Bruzual (1991).

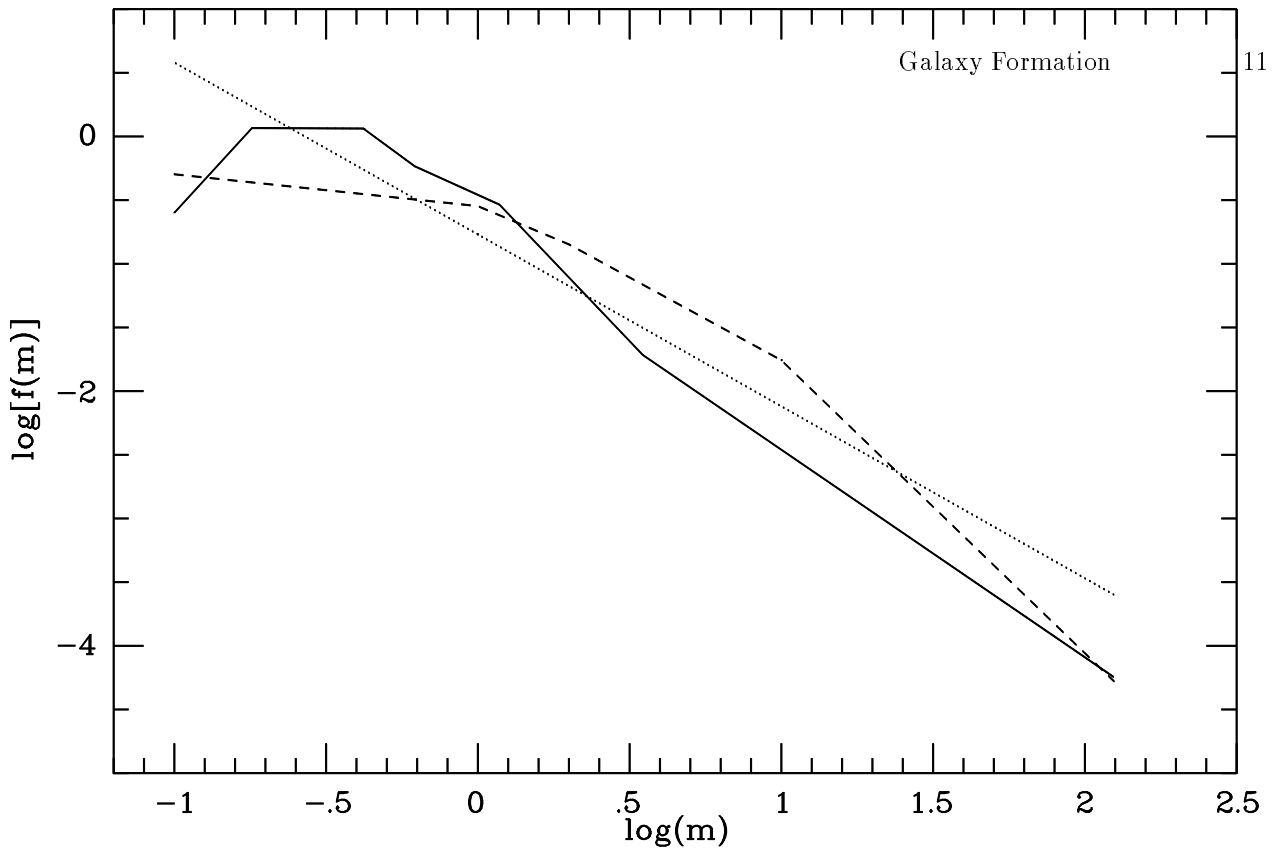

**Figure 4.**  Comparison of three IMFs. Solid line: Scalo (1986); dotted line: Salpeter (1955); dashed line: Miller & Scalo (1979). Here, $f(m)dm$ is the mass fraction in stars of mass $m$ to $m + dm$. Each IMF is approximated by a series of power-law segments $f(m) \propto m^{-x}$ (following the notation of Tinsley 1980 ) and normalized so that the total mass is $1 M_\odot$.

We use the IMF of the solar neighbourhood (Scalo 1986) as our fiducial choice. Figure 4 compares the Scalo (1986) IMF, with the Salpeter (1955) and Miller & Scalo (1979) IMFs in the interval $0.1 M_\odot < M < 125 M_\odot$. The mass locked in objects of mass $M < 0.1 M_\odot$ (brown dwarfs, planets, etc.) we will specify by the parameter, $\Upsilon$, which we define as the total mass in stars divided by the mass in luminous stars with mass greater than $0.1 M_\odot$. We allow ourselves the freedom to adjust this parameter in order to obtain a reasonable match to the bright end of the present day galaxy luminosity function. We will investigate the effect of changing both the shape and the upper mass cutoff of the IMF.

The output of Bruzual & Charlot's model is the integrated SED of the stellar population as a function of age, $t$ (normalized for $1 M_\odot$). From the SED of a single stellar population (instantaneous burst), $\Phi_\lambda(t)$, we compute the SED, $S_\lambda(t)$, of the composite stellar population of each of the galaxies in our model by performing the convolution integral

$$S_\lambda(t) = \int_0^t \Phi_\lambda(t - t') \, \dot{m}_\star(t') \, dt', \qquad (2.13)$$

where the star formation rate is given by equation (2.5). Using filter response curves, we compute broad-band observed magnitudes and colours at any redshift. We present results for Johnson $B$ and $K$, and Kron-Cousins $I$ bands (see, eg. Aragón-Salamanca et al. 1993 for details). Figure 5 shows some examples of the model results for a stellar population formed in one single instantaneous burst: the evolution with time of the absolute $M_B$ magnitude (normalized to $1 M_\odot$) and the $V - K$ and $B - V$ colours.

The models also provide the number of Lyman-continuum photons, which allows us to compute the luminosity of the H$\alpha$ emission line, for Case-B recombination, as a function of the total current SFR (see, eg., Kennicutt 1983). This line luminosity can then be used as a useful measure of the present star formation rate in external galaxies. In the absence of extinction, we find,

$$L(\mathrm{H}\alpha) = 9.40 \times 10^{40} \frac{\mathrm{SFR}}{M_\odot \, \mathrm{yr}^{-1}} \, \mathrm{ergs \, s}^{-1}, \qquad (2.14)$$

for the Scalo IMF with an upper mass cutoff of $125 M_\odot$. This result, obtained with a completely independent stellar evolution calculation, agrees with that of Kennicutt (1983) within 20%, the difference being mainly due to the use of a slightly different IMF. Extinction and [N II] contamination have not been included in the calculations because we will compare our results with observational data that have been corrected for these effects (see §4).

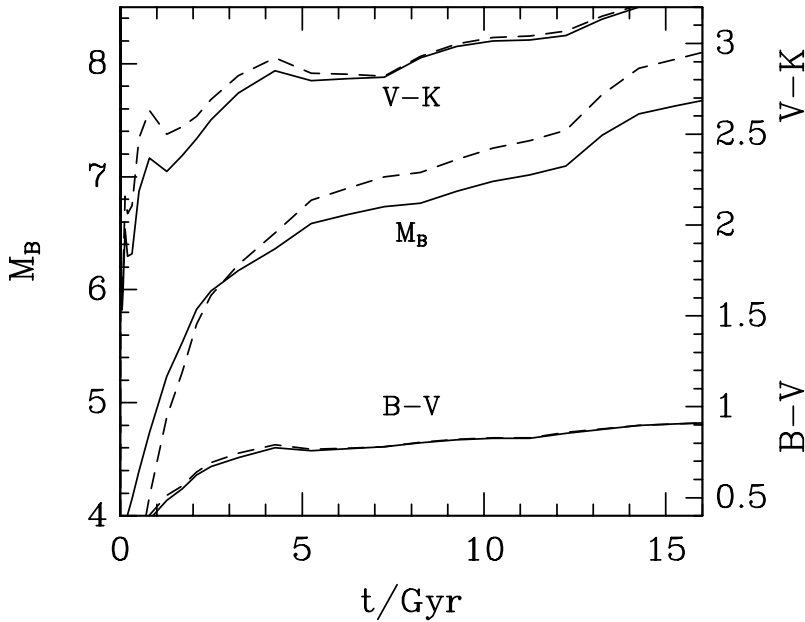

**Figure 5.**   The evolution with time of the absolute $M_B$ magnitude (normalized to $1 M_\odot$), and $B-V$ and $V-K$ colours for a single age stellar population. The solid lines show the evolution for the Scalo (1986) IMF and the dashed lines for the Miller & Scalo (1979) IMF, as given in figure 4.

## 3 Modelling Strategy

### 3.1 THE ALGORITHM FOR GALAXY FORMATION

We now describe how we graft together the ideas of the previous section to form one complete algorithm to follow the formation and evolution of a galaxy population in a hierarchical universe. The block model provides us with a description of the sequence of halo formation and merging and forms the framework in which we embed our approximate treatment of the physics of gas heating and cooling, star formation and galaxy merging.

It is worth recalling all the parameters which must be set in order to completely specify our model. The first category of parameters, $\Lambda_0$, $\Omega_0$, $\Omega_b$, $H_0$, and $\sigma_8$, define the cosmological model. With the exception of $\Omega_b$, we have chosen not to vary these in this study and, for illustration, we have adopted the standard parameters of the well known cold dark matter model. Next is the model of star formation and feedback. Here we have chosen to be guided by the numerical simulations of Navarro & White (1993). As described in §2.2, the one controlling parameter in these models is $f_v$, which quantifies the amount of energy from SN and evolving stars that is dumped as kinetic energy into the intragalactic gas. This, in turn, controls the star formation rate and the amount of gas that is ejected from a galaxy. The parameters $\alpha_\star$, $V_{\mathrm{hot}}$ and $\alpha_{\mathrm{hot}}$ have the values required to fit the results of these simulations. The zero point, $\tau_\star^0$, of the equation governing the star formation rate we treat as a free parameter. Next we must specify the IMF with which stars form. We adopt a standard, observationally determined, IMF for stars with masses in the range $0.1 < m/M_\odot < 125$ but, in addition, we assume that along with these visible stars, brown dwarfs are formed, so that the total mass in stars is $\Upsilon$ times larger than the mass in visible stars. Finally, the parameters $\tau_{\mathrm{mrg}}^0$ and $\alpha_{\mathrm{mrg}}$ which together determine the timescale for the merger of galaxies within one common DM halo, are also fixed by reference to the numerical simulations of Navarro, Frenk & White (1994).

The computational steps involved in evolving a realization of the block model are as follows:

i) We set all the model parameters including the output redshift.

ii) We create a realization of the block model, typically with total mass, $M_{\mathrm{block}} = 8 \times 10^{15} \mathrm{M}_\odot/\mathrm{h}$, and 20 levels of subdivision. We choose the density perturbation assigned to the top level of the block model in a systematic fashion from a gaussian of width $\sigma(M_{\mathrm{block}})$, so that a set of $n$ realizations gives a fair sample of both large scale underdense and overdense regions.

iii) Starting with the lowest level of the hierarchy, we select a halo which formed earlier than the required output redshift.

iv) We determine the physical parameters of this halo.

We compute its mass, $M_H$, from its position in the hierarchy and then, from its formation redshift, calculate the halo circular velocity, $V_c$, and virial temperature, $T_{\mathrm{vir}}$, using equations (2.2) and (2.3).

We compute the lifetime, $t_l$, of the halo. This is the elapsed time between the formation redshift of the halo and the earlier of either the redshift at which the halo is subsumed into a larger halo or the selected output redshift.

v) If the halo formed by the merger of other halos, all of which had masses below the resolution limit, (*i.e.* it is a halo in the lowest level of the hierarchy) then:

We set the mass of hot diffuse gas in the halo to $\Omega_b M_H$ and set the gas temperature to be the virial temperature, $T_{\mathrm{vir}}$, of the halo.

Otherwise, we loop over all the galaxies that find themselves within this new common halo as a result of the merger.

We merge the hot diffuse gas from each of these halos and, under the assumption that the shocks induced by the merger will efficiently heat the gas, we set its temperature to the virial temperature of the common halo.

Then, using equation (2.12) for each galaxy we compute the merger timescale, $\tau_{\mathrm{mrg}}$, within the new common halo

If $\tau_{\mathrm{mrg}} < t_l$,

we merge both the stars and cold gas reservoir from this galaxy into a single galaxy at the centre of the new halo. The circular velocity of this central, accreting, galaxy is set to the circular velocity of the common halo.

Alternatively if $\tau_{\mathrm{mrg}} > t_l$,

this galaxy remains as a satellite galaxy or group member with its original circular velocity, but without its own halo of hot gas. This has been stripped off and now forms part of the hot atmosphere bound to the common halo. We compute the mass of stars that form from this galaxy's remaining cold gas reservoir by integrating the exponential star formation law, (2.5), over the lifetime, $t_l$. The luminosity of these stars in any required band is computed by performing the convolution integral (2.13). We also compute, using (2.8) and (2.11), the mass of cold gas that is reheated as a result of the feedback process and transfer this material from the cold gas reservoir of this galaxy to the hot diffuse gaseous atmosphere of the common halo.

If none of the satellite galaxies satisfied the criterion for a merger, then the one with the largest circular velocity is selected to be the focus for the gas that later cools in the common halo. For this galaxy we retain its current value of $V_c$ rather than adopting that of the new halo.

vi) We now use the cooling curve to compute how much of the virialized hot gas in the common halo can cool in the lifetime $t_l$. This amount of gas is removed from the hot virialized component and put into the cold star forming reservoir of the central galaxy.

vii) We then compute the mass and luminosity of the stars that form from this new reservoir of cold gas and add these to the central galaxy.

viii) Now we move on to the next halo in this level of the hierarchy and repeat this procedure from step (iv)

ix) Having completed all these steps for one level of the hierarchy we move to the next level up in the hierarchy and repeat from step (iv) until the topmost level of the hierarchy is reached.

x) We then repeat the whole process again from step (ii) until we have made $n$ complete realizations which fairly sample a gaussian mix of underdense and overdense environments. Typically we use $n = 5$ realizations, corresponding to a total sample volume of $(52\mathrm{h}^{-1}\mathrm{Mpc})^3$, for $M_{\mathrm{block}} = 8 \times 10^{15} \mathrm{M}_\odot/h$

On completion of this cycle of computational steps we output a catalogue with an entry for each galaxy that exists at the chosen output time and, for each galaxy, we list its stellar mass, circular velocity, cold gas mass, and luminosity in each required observing band. The output for $z = 0$ can be used to compile statistical properties of the present galaxy population such as luminosity functions in various passbands, star formation rates as a function of luminosity, colour-magnitude distributions, etc. We use the outputs at higher redshifts to construct luminosity functions in the wavebands corresponding to the observed and rest frame $B$ and $K$ bands and use the former to predict the $B$- and $K$-band faint galaxy counts and redshift distributions.

Our current model does not incorporate chemical enrichment, but it would be quite straightforward to extend it to incorporate metal production. We could compute the yield for our selected IMF and this would then determine the metal production rate. We would then have to parameterize the fraction of these metals that go to enrich the cold star-forming gas and the remaining fraction that flows out of the galaxy as a hot SN driven wind, which then mixes with and enriches the halo gas. The enrichment of the halo gas would modify the cooling time of the gas, which we could easily compute. Equally important, the gradual enrichment of the star-forming gas would lead to the formation of stars with a range of metallicities. However, the stellar population synthesis code we are using assumes solar metallicity. For this code to be extended to follow the evolution of stellar populations of arbitrary metallicity would require theoretical isochrones and libraries of stellar spectra for a variety of metallicities. Supplied with the function $\Phi_\lambda(t)$, describing the evolution of the SED of a single-age population of stars, for a range of metallicities both, sub-solar and super-solar, this information would allow us to incorporate chemical enrichment within our models.

For simplicity, we have assumed that the halo gas is not enriched and remains at approximately primordial abundances during the epoch of galaxy formation, while the star forming gas undergoes prompt enrichment so that all stars form with approximately solar metallicity. This is clearly an oversimplification, but one which is forced upon us by the current state of development of stellar population synthesis models. The inclusion of metals would substantially increase the cooling rates, but this would not have a large effect as cooling is already very rapid in all but the largest galaxies. Chemical enrichment would also modify the colours of our galaxies. In particular, our current model may lack blue metal-poor galaxies at high redshift. Finally, we neglect the effects of dust on the visibility of our galaxies.

### 3.2 DATA

Our model calculates all the basic observable properties of galaxies. Thus there is a wealth of observational data that can be used to constrain the results. The data that we consider best for the purpose of defining the properties of present day galaxies are: the $B$- and $K$-band galaxy luminosity functions, the colour-magnitude distribution of bright galaxies, galaxy star formation rates as a function of luminosity and the $I$-band Tully-Fisher relation. Galaxy evolution at moderate redshift is then well constrained by faint $B$- and $K$-band galaxy counts, and their observed redshift distribution.

### 3.3 PRIORITIES

Our model of galaxy formation contains many poorly known quantities that parameterize the uncertain physics describing star formation, the details of the cosmological model and the rate of galaxy mergers. These various parameters are coupled, so the effects they have on the model interact with each other. Therefore they cannot be varied independently. Nevertheless, there is not infinite freedom within these models and the observational data place very interesting constraints on viable choices.

We have chosen to present our results by first comparing a fiducial model with all the observational data, pointing out where this model compares well with the data and where it is found lacking. We shall then vary each parameter individually and in combination, and note how the model predictions change by comparison to the fiducial model. In this way, we will learn which parameters are well constrained by the observational data and which remain relatively uncertain. We will also learn which failings of the fiducial model can be easily remedied by small changes to the input parameters and which require a drastic revision of the model.

## 4 Results

### 4.1 FIDUCIAL MODEL

In this section we make a thorough comparison of the properties of our fiducial model with observational data. Its successes and failures are summarized in §5. Listed below, and grouped according to the physics that they model, are the parameter values that define the fiducial model;

Cosmology

- Cold Dark Matter Power Spectrum
- $\Lambda_0 = 0$, $\Omega_0 = 1$, $H_0 = 50 \mathrm{kms}^{-1}/\mathrm{Mpc}$, and $\sigma_8 = 0.67$.
- $\Omega_b = 0.06$

Star Formation and Feedback

- $f_v = 0.2$ $\Rightarrow$ $\alpha_\star = -1.5$, $V_{\mathrm{hot}} = 140$ km s$^{-1}$ and $\alpha_{\mathrm{hot}} = 5.5$
- $\tau_\star^0 = 2.0\,\mathrm{Gyr}$

Stellar Population

- Scalo IMF and $\Upsilon = 2.7$.

Galaxy Merging

- $\tau_{\mathrm{mrg}}^0 = 0.5\,\tau_{\mathrm{dyn}}$ and $\alpha_{\mathrm{mrg}} = 0.25$.

Our fiducial model assumes the standard parameters of the CDM cosmology with a baryon fraction, $\Omega_b = 0.06$, compatible with the bounds placed by primordial nucleosynthesis (Walker *et al.* 1991). The other parameters specified above are reasonable and have been chosen primarily to produce a model that has an acceptable $B$-band luminosity function. The effect of varying these parameters will be considered in §4.2.

#### 4.1.1 Luminosity Functions

Figure 6 compares both the blue and infrared galaxy luminosity function of the fiducial model with observational data. The $B$-band data are taken from Loveday *et al.* (1992) and converted from their observed $b_j$ to Johnson $B$ assuming $B = b_j + 0.2$ (see Metcalfe *et al.* 1991). The $K$-band data are those of Mobasher *et al.* (1993).

Each model luminosity function has a power-law form at faint magnitudes and a break at the bright end. The existence of a break is the result of the cooling criterion of equation (2.4). Its actual position depends on the choice of IMF and, in particular, on the mass locked up in non-luminous stars. Thus, the characteristic luminosity, $L_\star$, is inversely proportional to our parameter $\Upsilon$. In this fiducial model, the value of $\Upsilon = 2.7$ was chosen as a compromise between producing a good fit to the bright, exponentially falling, end of the $B$-band luminosity function and simultaneously fitting the $K$-band luminosity function. We do not have unlimited freedom in our choice of $\Upsilon$ since it directly determines the stellar mass-to-light ratios of the galaxies and these are observationally constrained. Thus it can be considered a success that the value of $\Upsilon$ required to produce the observed $B$-band $L_\star$ implies a mean stellar mass-to-light ratio of $15h\mathrm{M}_\odot/\mathrm{L}_\odot$, which although on the high side is not inconsistent with observations.

The faint end slope of the $B$-band luminosity function is steeper than the estimate of Loveday *et al.* (1992). These authors find that the best fitting Schechter function has a faint end slope of $\alpha = 0.97 \pm 0.15$, while the faint slope in our model has $\alpha \approx 1.5$. Nevertheless, it is a significant improvement over the very steep, $\alpha \gtrsim 2$, functions found by WF, Cole (1991), and Kauffmann *et al.* (1993). The reason for this improvement is that we have adjusted the parameters controlling feedback (§2.2) and mergers (§2.3) to achieve a shallower luminosity function. (Some of the models of Lacey *et al.* (1993) with strong feedback also produce a relatively shallow faint end slope.) Having tuned the parameters in this way there is no further freedom in the model and so it is gratifying to see that we simultaneously achieve a good fit to the $K$-band luminosity function. It is also worth noting that estimates of the luminosity functions in clusters (*e.g.* Sandage *et al.* 1985, Impey *et al.* 1988) often find steeper faint end slopes than that of the field luminosity function plotted in Figure 6. An even better match to the two field luminosity functions would be achieved if the brightest galaxies were redder than in our fiducial model. This point is discussed in detail in the next section.

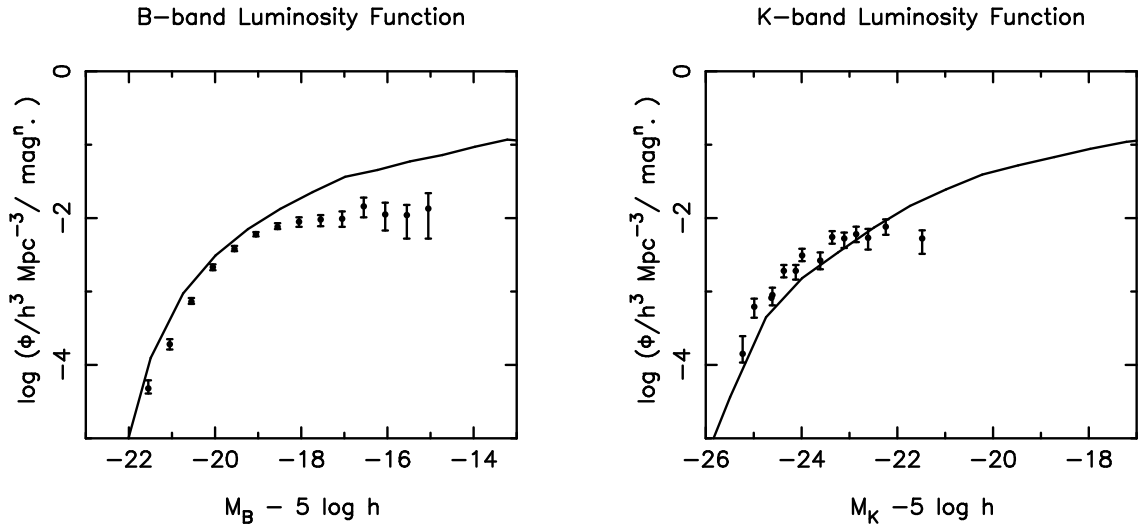

**Figure 6.** The luminosity functions in the $B$ and $K$-bands in the fiducial model. The $B$-band data displayed in (a) are the luminosity function estimated by Loveday *et al.* (1992) from the Stromlo-APM redshift survey. We have added 0.2 to their magnitudes to convert from their $b_J$ band to the standard Johnson $B$. The $K$-band data displayed in (b) show the infrared luminosity function estimated by Mobasher *et al.* (1993).

#### 4.1.2 Colours and Star Formation Rates

One of the primary challenges to models of galaxy formation is the broad colour distribution exhibited by galaxies in the local universe which reflects a wide variety of star formation histories. This broad range is clearly demonstrated in the histograms of $B - K$ colour shown in Figure 7. The spread in current star formation rates as measured by the H$\alpha$ luminosity [see §2.4 and equation (2.14)], is shown in Figure 8. It can be seen that with the exception of the faintest bin, which contains only 9 galaxies, the colour distribution and therefore the range of star formation histories is roughly independent of $B$ magnitude. This observation implies that galaxies with older, redder, stellar populations (ellipticals) are as bright in the $B$-band as galaxies with young stellar populations (spirals). Since in hierarchical scenarios the most massive galaxies generally form at more recent times, this observation poses a severe challenge which has not been successfully met by previous galaxy formation models of this type (see Lacey *et al.* 1993 and references therein). Simply put, hierarchical models have traditionally been unable to create luminous, red, elliptical galaxies.

A comparison of our fiducial model to observations of colours and H$\alpha$ luminosities of galaxies at the current epoch reveals some notable successes and a few failures. As can be seen in Figures 7 and 8, our fiducial model produces galaxies which span most of the observed range of $B - K$ colour and H$\alpha$ luminosity. Other colour indices show the same behaviour, with for instance $B - I$ colours spanning the range 1.6 to 2.1. This colour range indicates a true variety of star formation histories since our model does not include sources of scatter which exist in observations of real galaxies, such as metallicity variations, variations in extinction by dust, and observational error. More significantly, there is *no* trend for the model galaxies which are brighter in the $B$ band to have bluer colours than galaxies with fainter blue luminosities. We count this as a major success, particularly given the failure of other models in this regard. Note that had we included metallicities we would have likely found a trend for more massive galaxies to be more metal-rich and hence our models would have reproduced the well-known trend that brighter galaxies are slightly redder than fainter ones (e.g. Visvanathan & Sandage 1977, Mobasher *et al.* 1986).

Since our fiducial model is the first of its type to explicitly produce a colour-magnitude relationship in agreement with observation (sans metallicity effects), it is instructive to consider how this occurs in detail.

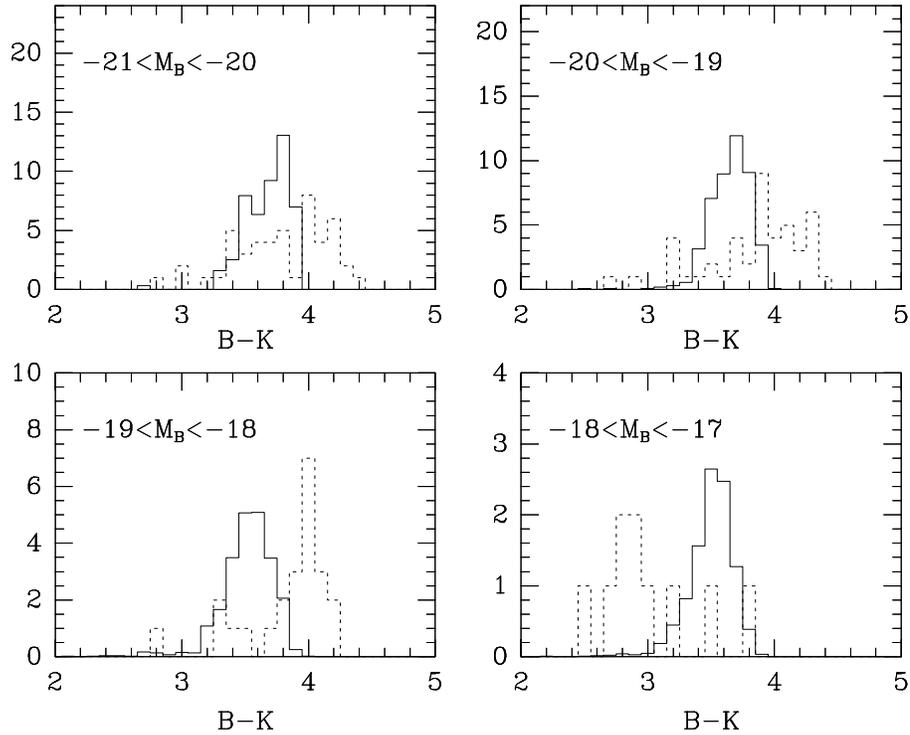

**Figure 7.** Histograms of the $B - K$ colour distribution for various values of the $B$-band absolute magnitude. The data (broken lines) are from Mobasher *et al.* (1986) and show the actual numbers of galaxies in the data set. The model (solid lines) has been normalized to enclose the same area as the data in each plot.

The two key processes are, firstly, that the objects which eventually merge to form many of the most massive galaxies form their stars early and, secondly, that these new larger galaxies fall into larger potential wells and thereby lose any remaining reservoir of gas which could have cooled and maintained star formation.

The primary weakness of the fiducial model regarding the properties of galaxies at the current epoch is that it fails to produce galaxies with colours as red as those of many normal elliptical galaxies. As Figure 7 shows, our reddest galaxies fall short by about 0.3 magnitudes in $B - K$. This is a major problem since, at face value, it implies that none of our galaxies are as old as the elliptical and lenticular galaxies which make up a fair fraction ($\sim 30\%$) of all of the galaxies observed in magnitude limited samples. It is natural to ask whether this discrepancy might be due to shortcomings in the understanding of the late stages of stellar evolution. It is important to note that the models reproduce the $B - K$ colours of elliptical galaxies approximately 14 Gyr after a burst of star formation. The red galaxies in our fiducial model have bluer colours because they are primarily composed of stars with ages of roughly 9 Gyr, and the stellar population models suggest that such objects are bluer than observed elliptical galaxies.

The discrepancy between the colours of the reddest galaxies in our fiducial model and those observed for early-type galaxies is not easily explained except in terms of ages. The age problem may be either in the ages of the oldest galaxies in our fiducial model or in the time it takes to produce red galaxies in the stellar population model, but other explanations do not appear to be viable. For example, although corrections for dust extinction are a significant source of uncertainty for the intrinsic colours of spiral galaxies, they are not believed to be important in ellipticals. The colour difference between the fiducial model and observations cannot be entirely explained by differences between the metallicity of the observed galaxies and the solar metallicity of the stellar spectra used in the stellar population model. Firstly, the difference in colour for the reddest galaxies persists over a range of magnitudes, including galaxies with luminosity near $L_\star$, which are observed to have approximately the metallicity assumed in the stellar population models. Secondly, the observed colours are integrated colours and so should be consistent with average metallicity close to solar, rather than with the supra-solar metallicities derived from spectra of the central regions of bright ellipticals.

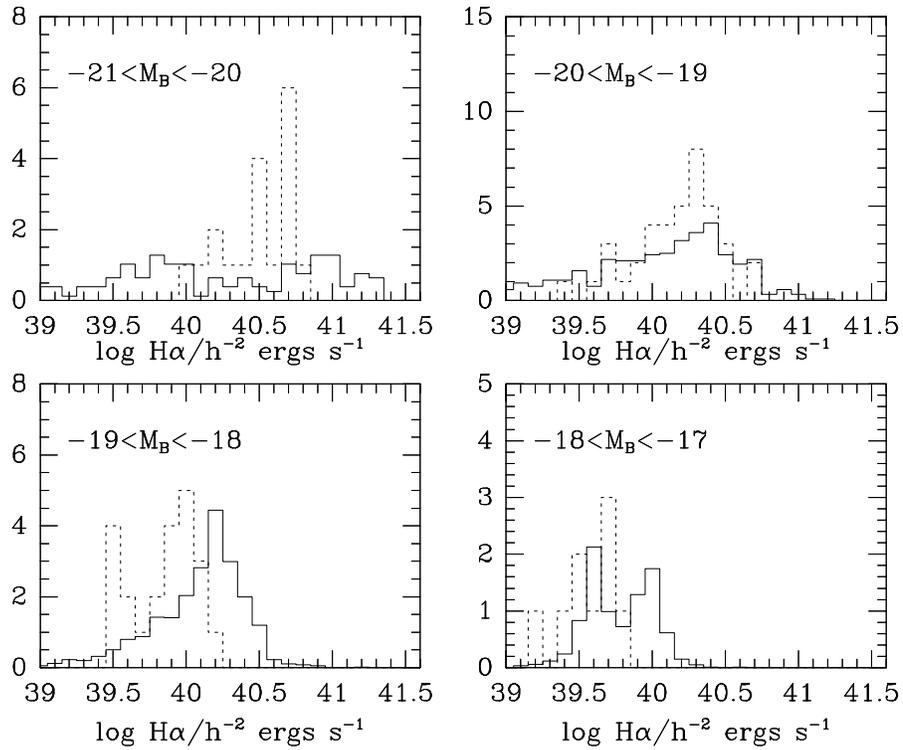

**Figure 8.** Histograms of the distribution of the H$\alpha$-line luminosity for star-forming galaxies at various fixed $B$-band absolute magnitudes. The data (broken lines) are from the Kennicutt (1983) survey of spiral galaxies. The model (solid lines) is in general agreement with the data. The lower limit of the plot at $10^{39}$ h$^{-2}$ ergs/s is roughly the lower limit of the observations. Only an upper limit to the H$\alpha$ flux was reported for most of the Sa galaxies and a few of the Sab galaxies observed by Kennicutt. The H$\alpha$ fluxes of elliptical and S0 galaxies, which were not included in the Kennicutt sample, are also normally below the limit of this plot. In total, the fraction of galaxies in these magnitude ranges without detectable H$\alpha$ emission at the approximate limit of the Kennicutt survey ($10^{39}$ h$^{-2}$ ergs/s) ranges from about 40% in the brightest bin to about 20% in the faintest bin. For comparison, the fraction of model galaxies with H$\alpha$ fluxes lower than $10^{39}$ h$^{-2}$ ergs/s is 22% for the brightest bin, 7.4% for galaxies with $-20 \leq M_B < -19$, 0.6% for galaxies with $-19 \leq M_B < -18$, and less than 0.1% for galaxies in the faintest bin.

### 4.1.3 Faint Galaxy Number Counts

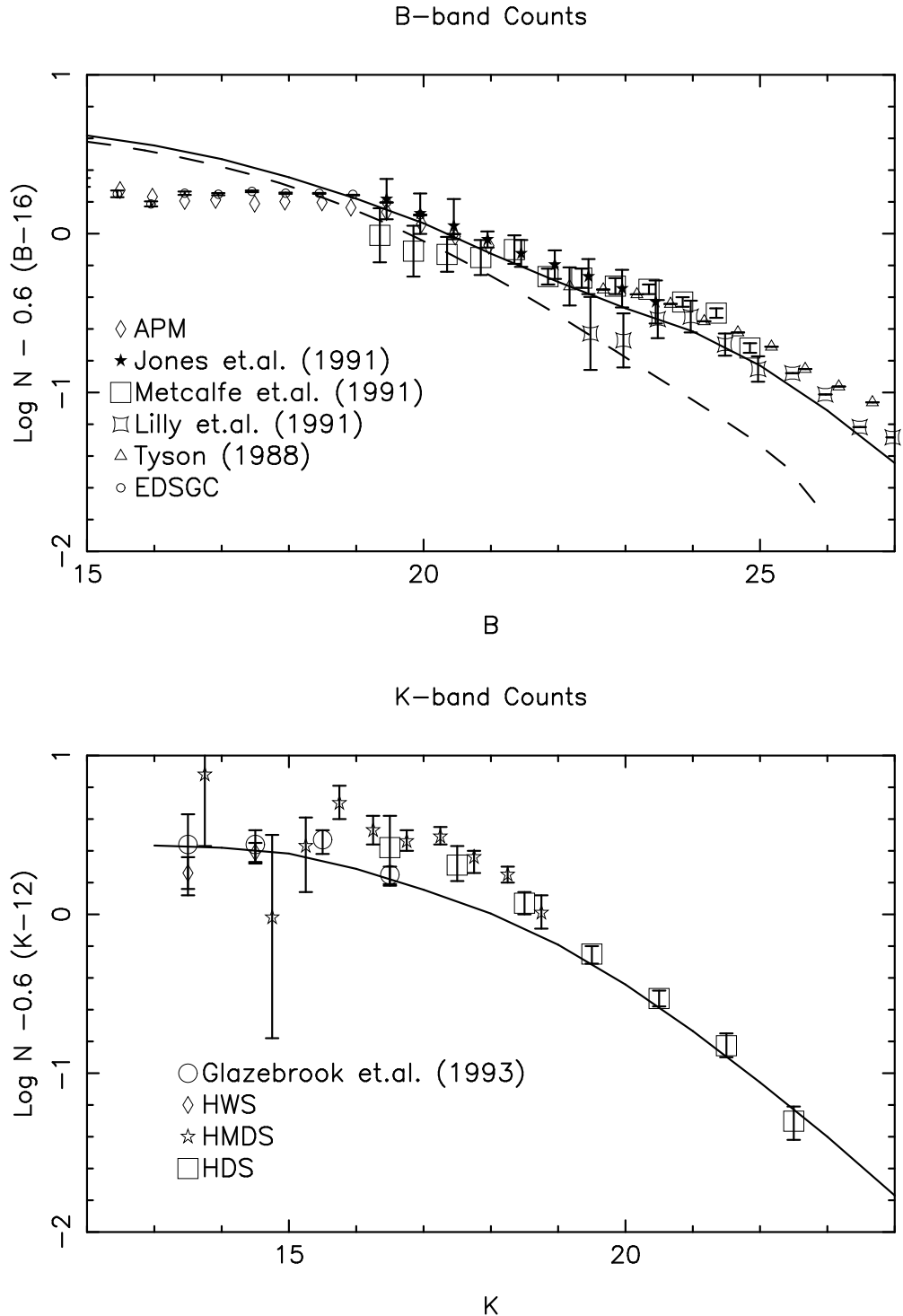

Figure 9. Differential galaxy number counts, N, per magnitude and per square degree as a function of apparent magnitude in the $B$ and $K$ bands. The raw counts have been divided by a pure power law with slope 0.6, so as to expand the useful dynamic range of the figure. Thus, Euclidean number counts would appear as a horizontal line. In each figure the number counts produced by the fiducial model are shown by the solid curve. The dashed curve is the no evolution model of Metcalfe et al. (1993). The $B$-band data in (a) are taken from Maddox et al. (1990b) (APM), Jones et al. (1991), Metcalfe et al. (1991), Lilly et al. (1991), Tyson (1988) and Heydon-Dumbleton et al. (1989)(EDSGC). Where necessary $b_j$ magnitudes have been converted to Johnson $B$ assuming $B = b_j + 0.2$. The $K$-band data in (b) are taken from Glazebrook et al. (in preparation), the Hawaii Wide Survey (HWS), the Hawaii Medium Deep Survey (HMDS) and the Hawaii Deep Survey (HDS) as reported by Gardner et al. (1993).

Traditionally, faint counts have been used to constrain the evolution of galaxies at moderate redshift. Figure 9 shows blue and infrared faint galaxy number counts. The plots show the raw number counts after division by a power-law with the Euclidean slope of 0.6. This has the desirable effect of compressing the ordinate of the figure, thereby increasing the dynamic range and emphasising the differences between the predictions and the data.

The bright APM and EDSGC galaxy counts are based on automatic scans of Schmidt photographic plates and are therefore subject to different sources of systematic error than the deeper counts estimated from CCD images. The near Euclidean slope found for the bright counts ($B < 18$) is impossible to reproduce in any smoothly varying model of galaxy evolution. Thus, as can be seen in Fig. 9, both our fiducial model and a no-evolution model, shown as the dashed curve (Metcalfe *et al.* 1993), overproduce the bright galaxy counts. In our model this excess is directly attributable to the shape of the B-band luminosity function which has more faint galaxies, *ie* a steeper faint end slope, than the Loveday *et al.* (1992) estimate from the APM-Stromlo redshift survey.

Overall the faint galaxy number counts predicted by the fiducial model are in remarkably good agreement with both the observed $B$ and $K$-band counts. The predicted $B$-band counts faintwards of $B = 23$ are slightly low in comparison to the observed counts. However our model is considerably closer to the observations than a simple no-evolution model. For example, at $B = 25$ the no-evolution model is a factor 4 below the observed counts while our fiducial model is only 36% too low. We shall see in §4.2.4 that, in fact, the $B$-band counts are quite sensitive to details of the shape of the assumed IMF.

### 4.1.4 Faint Galaxy Redshift Distribution

A strong constraint on galaxy evolution models that seek to explain the $B$-band counts is their shallow observed redshift distribution. Models that invoke luminosity evolution to produce the steep $B$-band counts invariably predict a high redshift tail in the galaxy distribution. In Figure 10 we compare our fiducial model with the recent data of Colless *et al.* (1993), who now have a 95% complete redshift sample of galaxies with blue magnitudes in the range $21 < b_j < 22.5$.

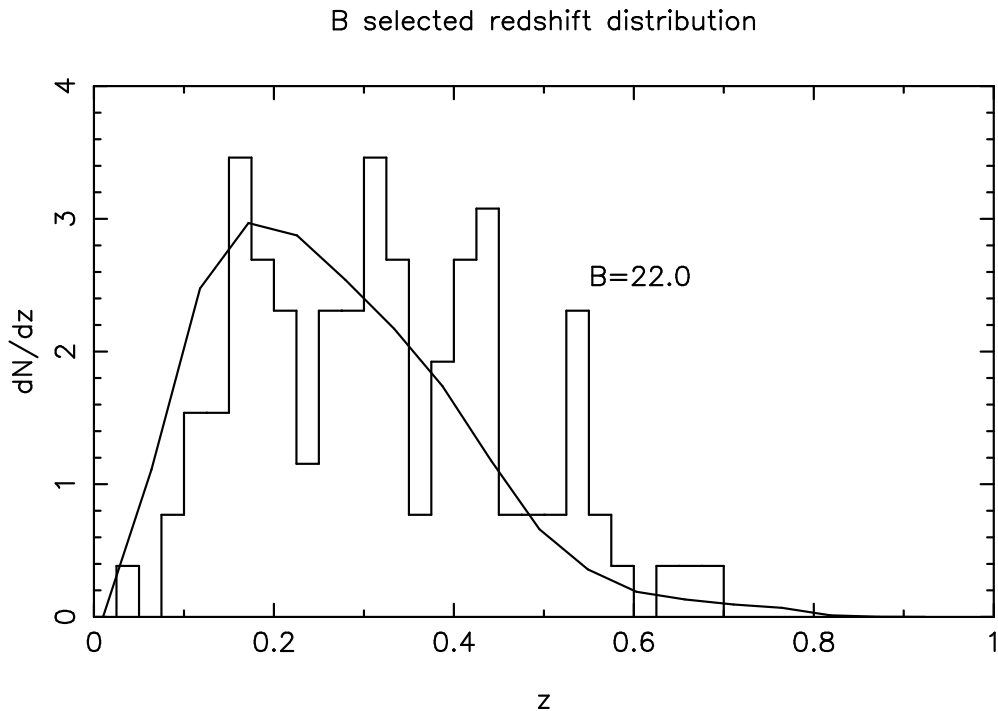

**Figure 10.** The redshift distribution of galaxies selected at an apparent magnitude of $B = 22$, compared to the observed redshift distribution of Colless *et al.* (1993). The histogram is the observed redshift distribution and the solid curve the prediction of our fiducial model. The two distributions have been normalized to enclose equal areas.

Compared to the observed distribution the fiducial model predicts a slight excess of galaxies at very low redshift. This feature is directly related to the faint end slope of the galaxy luminosity function, which is steeper for this model than observed. The fiducial model does not produce a tail of high redshift galaxies and has a low median redshift, as low or lower than the observed distribution. Thus, the reason why it produces higher counts than a no-evolution model is principally galaxy merging rather than luminosity evolution. This conclusion is reinforced in §4.3.1 where we present the predicted evolution of the galaxy luminosity function.

#### 4.1.5 The Tully-Fisher Relation

We now turn to the Tully-Fisher correlation between galaxy luminosity and circular velocity. This correlation is of a distinctly different nature to the other statistical properties of the present day galaxy population that we have considered so far. In this case, we compare the luminosity of a galaxy with a dynamical property, $V_c$, which, in our model, is determined purely by the DM halo in which the galaxy formed. The observed Tully-Fisher relation applies to undisturbed spiral galaxies. We have made no attempt to morphologically classify the galaxies in our model which must therefore be interpreted as containing an admixture of spirals, ellipticals and irregulars. Despite this complication the comparison proves to be very interesting.

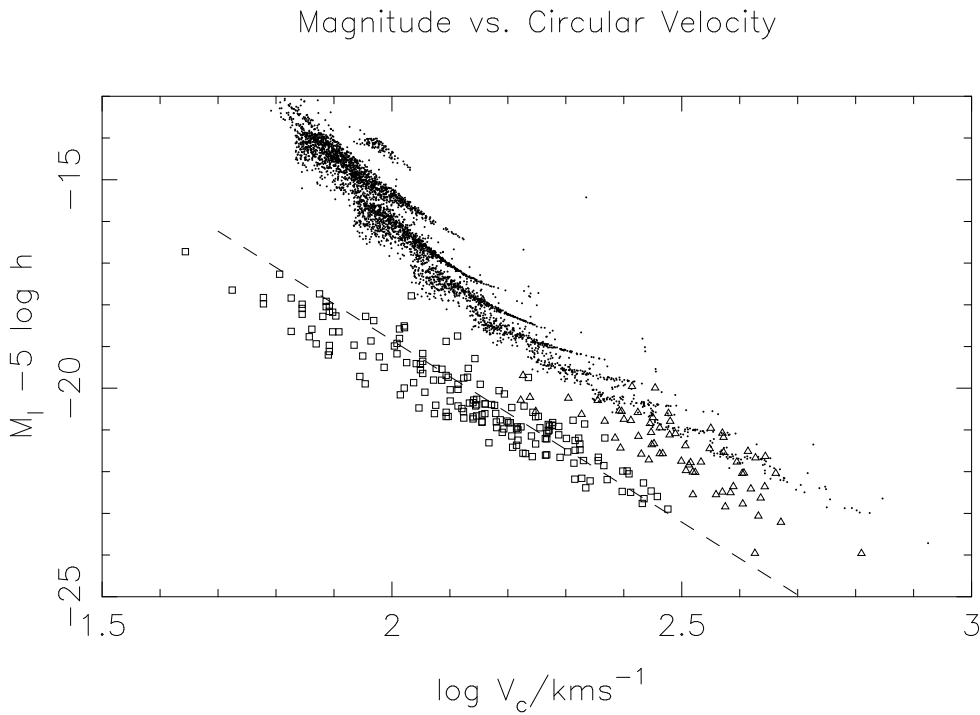

**Figure 11.** The infrared Tully-Fisher relation between I-band magnitude and circular velocity. The dashed line shows the locus of the best fit Tully-Fisher relation as determined by Pierce & Tully (1992) scaled to a Hubble constant, $H_0 = 100 \mathrm{kms}^{-1}/\mathrm{Mpc}$. The open squares are a sample of spirals compiled from new and published cluster data (Young *et al.* 1993 in preparation) and the triangles a sample of ellipticals from the Coma cluster (Lucey *et al.* 1991) which have been placed on this plane by defining an effective circular velocity in terms of the observed velocity dispersion, $V_c = \sqrt{3}\sigma_{1D}/1.1$. The points are the galaxies of the fiducial model.

Figure 11 compares the infrared Tully-Fisher relation in our fiducial model with observational data. The dashed line shows the locus of the best fit Tully-Fisher relation as determined by Pierce & Tully (1992). The square and triangular symbols give the locii of a sample of spiral galaxies compiled from new and published cluster data (Young *et al.* 1993 in preparation) and a sample of elliptical galaxies from the Coma cluster (Lucey *et al.* 1991), respectively. The Tully-Fisher relation of the model is offset from the observed relation. However the slope of and the scatter about the mean relation are in reasonable agreement with the observations, although it should be noted that the scatter in the observed correlation includes instrumental scatter. At $V_c = 200 \mathrm{\,km\,s}^{-1}$, the fiducial model gives $L \propto V_c^{3.6}$ and a scatter of 0.5 magnitudes compared to the slope and scatter determined by Pierce and Tully of 3.5 and 0.3 magnitudes respectively. This is very encouraging: simple models that incorporate cooling in each generation of halos but ignore feedback and assume all cold gas is converted into stellar populations with the same mass-to-light ratio predict a shallower slope, $L \propto V_c^2$, and scatter of $\sim 1$ magnitude (Cole & Kaiser 1989).

However, the "Tully-Fisher" relation for the model is significantly offset from the observed relation. The locus of elliptical galaxies, (which have been placed on this plot by assuming a mean $V - I = 1.17$ and an effective circular velocity given by $V_c = \sqrt{3}\sigma_{1D}/1.1$, as in Frenk et al. 1988), overlaps that of the model, but the locus of the more numerous spiral galaxies is quite distinct. The offset from the spiral correlation can be viewed as a vertical displacement on Figure 11, in which case the model galaxies are 1.8 magnitudes too faint, or as a horizontal displacement, in which case the model circular velocities are 60% too large. The latter interpretation is more telling since the galaxy luminosities have already been normalized to the bright end of the observed galaxy luminosity function. The offset in the Tully-Fisher relation is a very robust prediction of our model and cannot be significantly reduced by altering the model feedback parameter, merger parameters, or IMF.

The circular velocities that we have assigned to our model galaxies are the circular velocities of the DM halos in which they formed. The scalelength of the galaxy is, of course, much smaller than that of its surrounding DM halo; thus in equating these two circular velocities we are implicitly assuming that the combined halo/galaxy circular velocity curve is flat over a wide range of scales. This may not be a good approximation. Blumenthal et al. (1986), Barnes (1986), and Navarro & White (1993b) have shown that dissipation may, in fact, concentrate the baryonic component by a significant factor and so boost the galactic circular velocity above that of the surrounding halo (see also Persic & Salucci 1990). However, this effect goes in the wrong direction to help eliminate the offset in our model. We can think of no plausible systematic effect which would lower the galactic circular velocity below that of its surrounding halo, although some such systematic effect remains a possibility.

The origin of the offset can be understood in terms of the number density of halos, $\phi_H(V_c)dV_c$, as a function of circular velocity. This is fixed by the CDM power spectrum and our choice of normalization, $\sigma_8$. The luminosity, $L^\star$, of a typical galaxy, defined by the break in the observed galaxy luminosity function, can be related to a typical circular velocity, $V^\star$, using the observed Tully-Fisher relation. If the model were consistent with the data the number density of halos of circular velocity $V^\star$ should approximately equal the number density of galaxies of luminosity $L^\star$: $\phi_H(V^\star) \approx \phi(L^\star)$. However, for the CDM model we actually find $\phi_H(V^\star) > \phi(L^\star)$. Around $V^\star$, $\phi_H(V_c)$ is a decreasing function of $V_c$. Thus, to match the galaxy number density in our model to the observed luminosity function, we are forced to adopt a mass-to-light ratio such that the galaxies with a given $V_c$ end up being too faint. Possible solutions to this problem may require altering the cosmological model, for example by lowering the value of $\Omega_0$, which directly reduces the number density of halos, or by having a power spectrum with less small scale power such as in the mixed (hot and cold) dark matter model (Davis et al. 1992; Taylor & Rowan-Robinson 1992; Klypin et al. 1993) so that fewer halos with circular velocities around $V^\star$ are produced.

### 4.2 SENSITIVITY TO MODEL PARAMETERS

In the following subsections we consider the effects of varying the model parameters away from the values of the fiducial model. For each parameter variation we highlight the model predictions that are appreciably changed while noting those that are little affected. This exercise is particularly useful as it teaches us which observations constrain which physical processes.

#### 4.2.1 Feedback and Mergers

The most important parameters in our model are $f_v$, which determines the strength of the feedback process whereby SN energy regulates further star formation, and $\tau_{\rm mrg}^0$ and $\alpha_{\rm mrg}$, which together determine the frequency of galaxy mergers. The parameter $\alpha_{\rm mrg}$ has only a minor influence as the frequency of galaxy mergers is primarily determined by $\tau_{\rm mrg}^0$ so throughout this subsection we adopt $\alpha_{\rm mrg} = 0.25$. Increasing $f_v$ increases the fraction of gas that is expelled from galaxies forming in shallow potential wells, i.e. halos with low circular velocities. Decreasing $\tau_{\rm mrg}^0$ increases the merger frequency of galaxies in groups and clusters. These two processes have a strong influence on the shape of the galaxy luminosity function and consequently on the faint galaxy number counts and redshift distribution. It is necessary to vary both parameters simultaneously since their effects are strongly coupled. To illustrate this we concentrate on the behaviour of the $B$-band luminosity function; the variation of the $K$-band luminosity function is similar.

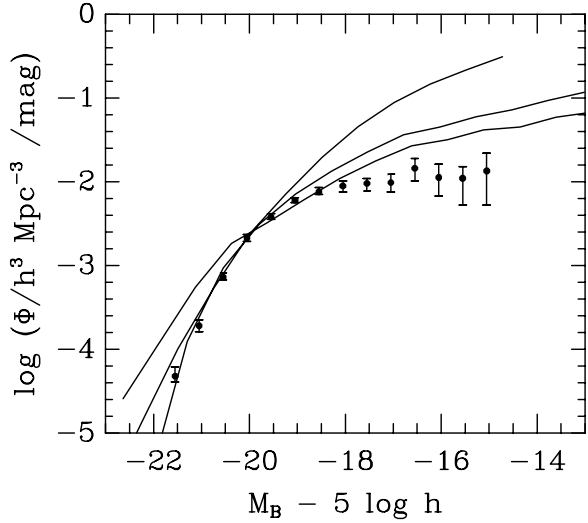

**Figure 12.** The dependence of the predicted $B$-band luminosity function on the timescale, $\tau^0_{\mathrm{mrg}}$, controlling the frequency of galaxy mergers. The "feedback" parameter is set to its fiducial value, $f_v = 0.2$. In order of decreasing steepness, the curves correspond to $\tau^0_{\mathrm{mrg}} \gg \tau_{\mathrm{dyn}}$, $\tau^0_{\mathrm{mrg}} = 0.5\tau_{\mathrm{dyn}}$ and $\tau^0_{\mathrm{mrg}} = 0.1\tau_{\mathrm{dyn}}$. The corresponding values adopted for the parameter $\Upsilon$, controlling the stellar mass-to-light ratio, are 1.1, 3.2 and 3.15. These values have been chosen to bring the luminosity functions into agreement with the observed luminosity function at $M_B - 5\log h \approx -20$. Increased merger rates flatten the slope of the faint end of the luminosity function, but very high rates lead to the production of a power law tail of excessively bright galaxies.

Figure 12 shows the change in the shape of the present day $B$-band luminosity function as the merger timescale, $\tau^0_{\mathrm{mrg}}$, is varied, while the feedback parameter is kept at its fiducial value of $f_v = 0.2$. The data displayed in this and the two subsequent figures are the $B$-band luminosity function estimated by Loveday et al. (1992). The case of no galaxy mergers, $\tau^0_{\mathrm{mrg}} \gg \tau_{\mathrm{dyn}}$, produces the luminosity function with the steepest faint end slope and the fewest very bright galaxies. Decreasing $\tau^0_{\mathrm{mrg}}$ flattens the luminosity function by causing faint galaxies to merge with each other and with bright galaxies. The curve of intermediate slope corresponds to our fiducial value, $\tau^0_{\mathrm{mrg}} = 0.5\tau_{\mathrm{dyn}}$. A shorter merger timescale, $\tau^0_{\mathrm{mrg}} = 0.1\tau_{\mathrm{dyn}}$, reduces further the number of faint galaxies, at the expense of creating, by mergers, a few excessively bright galaxies which give a power-law tail to the luminosity function at the bright end.

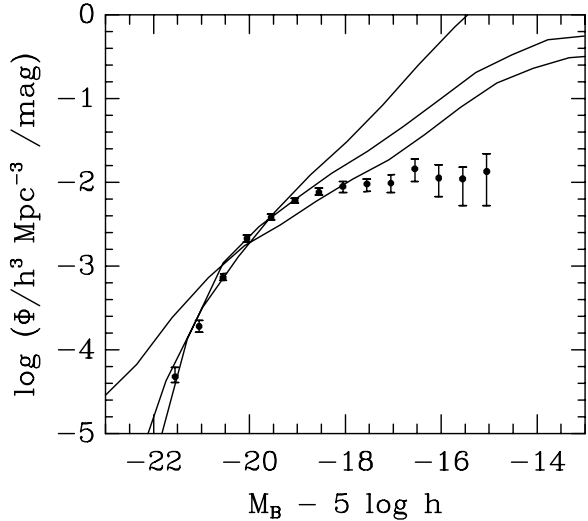

**Figure 13.** The dependence of the predicted $B$-band luminosity function on the timescale, $\tau^0_{\mathrm{mrg}}$, controlling the frequency of galaxy mergers, for the case of very little stellar feedback, $f_v = 0.01$. The three curves correspond to the same values of $\tau^0_{\mathrm{mrg}}$ used in Figure 12. The faint end of the luminosity function is steeper than the corresponding curves from Figure 12 in which $f_v = 0.2$. Again, an increased merger rate helps to flatten the slope of the faint end of the luminosity function, but very high rates lead to the production of a power-law tail of excessively bright galaxies. As in Figure 12, the parameter, $\Upsilon$, controlling the stellar mass-to-light ratio, has been adjusted in each case to bring the luminosity functions into agreement with the observed luminosity function at $M_B - 5\log h \approx -20$. In order of decreasing merger timescale the values of $\Upsilon$ required are 1.0, 3.2 and 4.0.

Figure 13 shows the change in the shape of the present day $B$-band luminosity function as the merger timescale, $\tau^0_{\rm mrg}$, is varied, and the strength of stellar feedback is reduced by setting $f_v = 0.01$. For each value of $\tau^0_{\rm mrg}$, the faint end slope of the luminosity function is steeper than the corresponding curve from Figure 12. As before, increasing the merger rate does flatten the slope of the luminosity function, but since feedback has not inhibited the formation of low luminosity galaxies the effect is not as strong as in Figure 12. Once again, at the bright end of the luminosity function, a high frequency of mergers, $\tau^0_{\rm mrg} = 0.1\tau_{\rm dyn}$, has a detrimental effect, giving rise to a power law tail of bright galaxies.

Finally, Figure 14 shows how for the moderate merger rate of the fiducial model, $\tau^0_{\rm mrg} = 0.5\tau_{\rm dyn}$, the faint end of the luminosity function steepens as $f_v$ is reduced. The curve with the shallowest slope corresponds to $f_v = 0.2$ while the steeper curves are for $f_v = 0.1$ and $0.01$. A luminosity function with a sufficiently shallow faint end slope extending brightwards all the way to the bright exponential cutoff can only be achieved by a combination of strong feedback and some merging of faint galaxies.

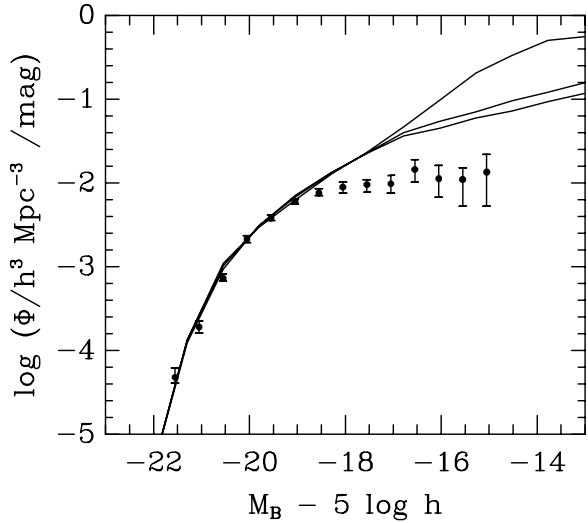

**Figure 14.** The dependence of the predicted $B$-band luminosity function on the parameter, $f_v$, controlling the strength of stellar feedback, for the fiducial merging timescale, $\tau^0_{\rm mrg} = 0.5\tau_{\rm dyn}$. In order of decreasing steepness, the curves correspond to $f_v = 0.01$, $0.1$, and $0.2$ respectively. Increasing $f_v$ inhibits star formation in low mass galaxies and, with the aid of a moderate galaxy merger rate, effectively flattens the faint end of the luminosity function, without producing an unwanted tail of excessively bright galaxies. The parameter, $\Upsilon$, controlling the stellar mass-to-light ratio was kept fixed at $\Upsilon = 3.2$ in these three models.

Changing the shape of the galaxy luminosity function has a knock-on effect on the faint galaxy counts and their redshift distribution. A steeper faint end slope produces an increased contribution to the galaxy counts from intrinsically faint nearby galaxies. This slightly increases the predicted counts and skews towards shallower redshifts the distribution of galaxies selected at an apparent magnitude of $B = 22$. Apart from this effect these parameter changes have very little influence on any of the other statistics considered in §4.1.

### 4.2.2 Star Formation Timescale

In the fiducial model and its variants considered above, the star formation timescale was kept fixed at $\tau^0_\star = 2.0\,{\rm Gyr}$. We now consider the effect of varying $\tau^0_\star$, while retaining the dependence on $V_c$ given by equation (2.10). Thus, we shall keep the relative star forming efficiency implied by the feedback model with $f_v = 0.2$, and only investigate the effect of globally increasing or decreasing the star formation timescale.

Altering the star formation timescale affects mainly the mean and scatter of current star formation rates and the colours of present day galaxies. Adopting a larger value of $\tau^0_\star$ suppresses star formation and the associated reheating of gas by young stars and SN at early times when the age of the universe was comparable to or smaller than $\tau^0_\star$. As a result, at the present day there are fewer old red stars and more cold gas to fuel current star formation. Figure 15 compares the distribution of galaxy colours produced for $\tau^0_\star = 4.0{\rm Gyr}$ with those of our fiducial value, $\tau^0_\star = 2.0{\rm Gyr}$. For the longer star formation timescale, the mean colour is bluer by approximately 0.3 magnitudes. The effect on the current star formation rates can be seen in Figure 16. The low value of $\tau^0_\star$ in the fiducial model produces both galaxies with high current star formation rates and galaxies which have nearly exhausted their reservoirs of cold gas and therefore have very low current star formation rates. With the larger value of $\tau^0_\star$, the median star formation rate is slightly greater and the scatter is much reduced. For both values of $\tau^0_\star$ the total mass of stars formed and their $B$-band mass-to-light ratio are much the same. As a result, changing $\tau_\star^0$ in this range has very little effect on any of the other galaxy properties that we have studied: the luminosity function, galaxy number counts, or the Tully-Fisher relation. If $\tau_\star^0$ were increased even further to the point where it became comparable or larger than the present age of the universe then, relative to the fiducial model, star formation would be supressed at all epochs and less stars would be formed overall.

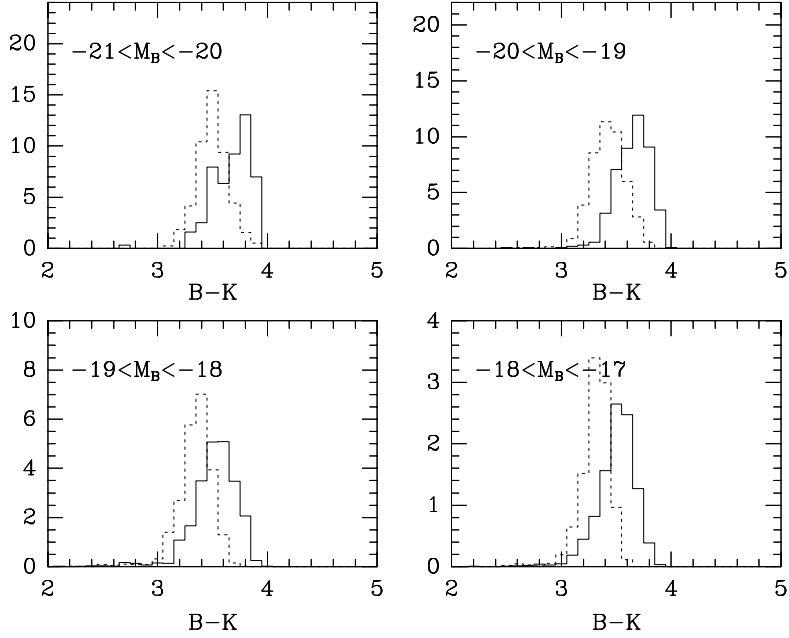

**Figure 15.** Histograms of the $B - K$ colours at various fixed $B$-band absolute magnitudes. The solid lines show the distributions for the fiducial model with star formation timescale, $\tau_\star^0 = 2.0$Gyr, while the broken lines show the distributions corresponding to $\tau_\star^0 = 4.0$Gyr.

Finally, we tried decreasing $\tau_\star^0$ below 2Gyr to see if this would shift the colour distribution even further to the red and provide a better match to the observed colour distribution. However, galaxies do not become significantly redder for values of $\tau_\star^0 \leq 2$Gyr. In this regime, the initial epoch of star formation is determined by the time at which the first potential wells are assembled in which efficient star formation can occur and not by the ratio of the star formation timescale to the age of the universe.

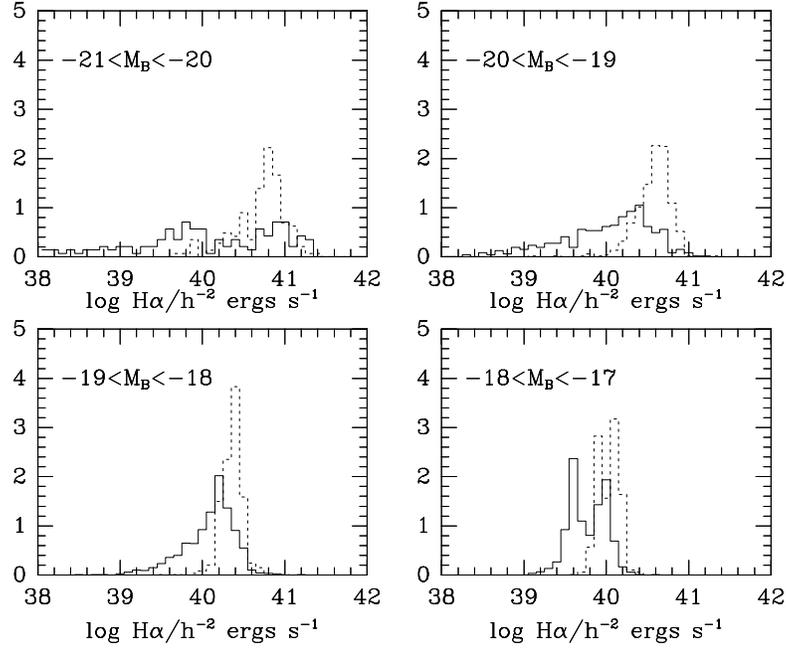

**Figure 16.** Histograms of the Hα-line luminosity for star-forming galaxies at various fixed $B$-band absolute magnitudes. The solid lines show the distributions for the fiducial model with star formation timescale, $\tau_\star^0 = 2.0$Gyr, while the boken lines show the distributions corresponding to $\tau_\star^0 = 4.0$Gyr.

### 4.2.3 Baryon Fraction

Increasing the assumed baryon fraction, $\Omega_b$, affects both the gas cooling rates through equation (2.4) and the amount of material available for star formation. We considered the effect of changing the baryon fraction from $\Omega_b = 0.06$ which, for $H_0 = 50\,{\rm km\,s^{-1}Mpc^{-1}}$, is the upper bound allowed by nucleosynthesis considerations (Walker *et al.* 1991), to $\Omega_b = 0.2$ which, again for $H_0 = 50\,{\rm km\,s^{-1}Mpc^{-1}}$, is close to the baryon fraction inferred by White *et al.* (1993b) from an inventory of the Coma cluster. Such a large change increases the total mass of stars formed by a large factor and, if $\Upsilon$ were left unaltered, would increase the luminosity of each galaxy by a similar factor. The mass fraction in non-luminous stars can be increased by increasing $\Upsilon$. This results in very little net change to the model luminosity function but at the expense of requiring stellar mass-to-light ratios of $80h{\rm M}_\odot/{\rm L}_\odot$ which are excluded by observations. The excessively large mass that is turned into stars could be reduced by greatly increasing $\tau_\star^0$ so reducing the number of stars formed, but this has the unwanted side-effect of making the mean galaxy colours excessively blue.

A more modest increase in $\Omega_b$ is permissible and, in this case, the effects on the model predictions are much the same as decreasing the star formation timescale as discussed above. Thus, increasing $\Omega_b$ leads to a larger spread in galaxy colours and star formation rates.

### 4.2.4 Initial Mass Function

The stellar initial mass function (IMF) determines the spectral evolution of the stellar populations that are formed. Although the IMF is reasonably well determined in the solar neighbourhood (Scalo 1986) there is little evidence that it is universal. It is therefore important to identify which aspects of our model are sensitive to the choice of IMF.

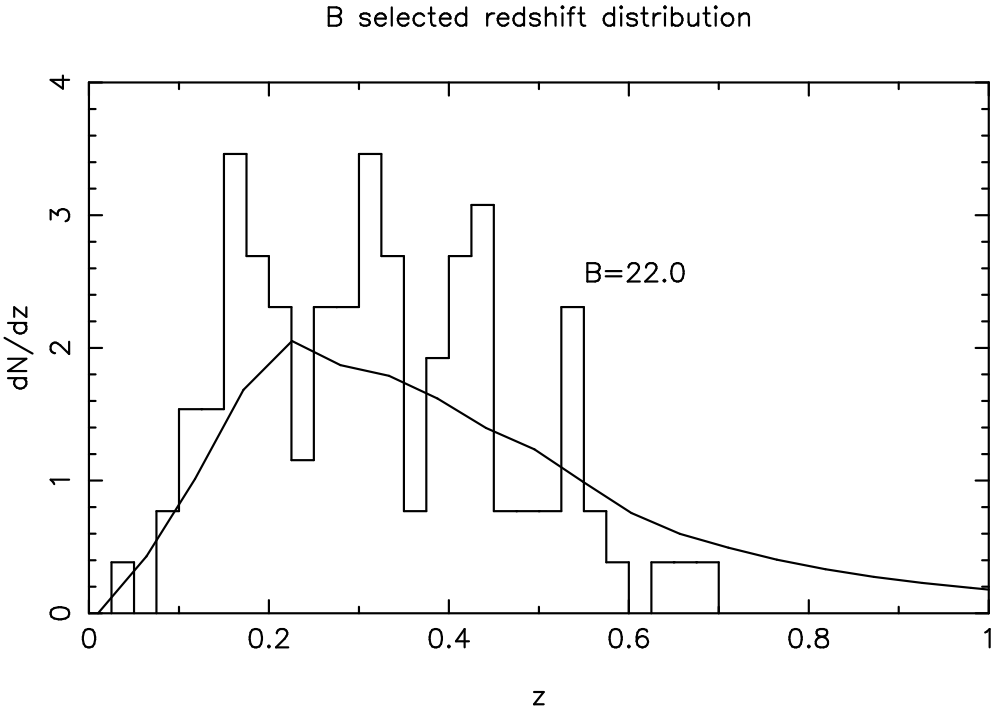

**Figure 17.** The predicted redshift distribution of galaxies selected at an apparent magnitude of $B = 22$ compared to the observed redshift distribution of Colless *et al.* (1993). The solid line shows the prediction for a model with the same parameters as the fiducial model except that the stars were assumed to form with the Miller-Scalo rather than the Scalo IMF (see Figure 4).

We first consider varying the high mass cutoff but otherwise retaining the form of the Scalo IMF (Figure 4). As the high mass cutoff is reduced from $125{\rm M}_\odot$ to $30{\rm M}_\odot$, we find that, with the exception of the H$\alpha$ flux, there is no noticeable change in any of the statistical properties studied in §4.1. The H$\alpha$ flux, used to measure the star formation rate, is quite sensitive to the high mass cutoff in the IMF because the majority of Lyman continuum photons that are degraded to produce H$\alpha$ are produced in very massive stars. Thus, a poor choice of the cutoff mass could lead to systematically high or low H$\alpha$ fluxes, while variations of the cutoff from galaxy to galaxy could lead to an increased scatter in the inferred star formation rates.

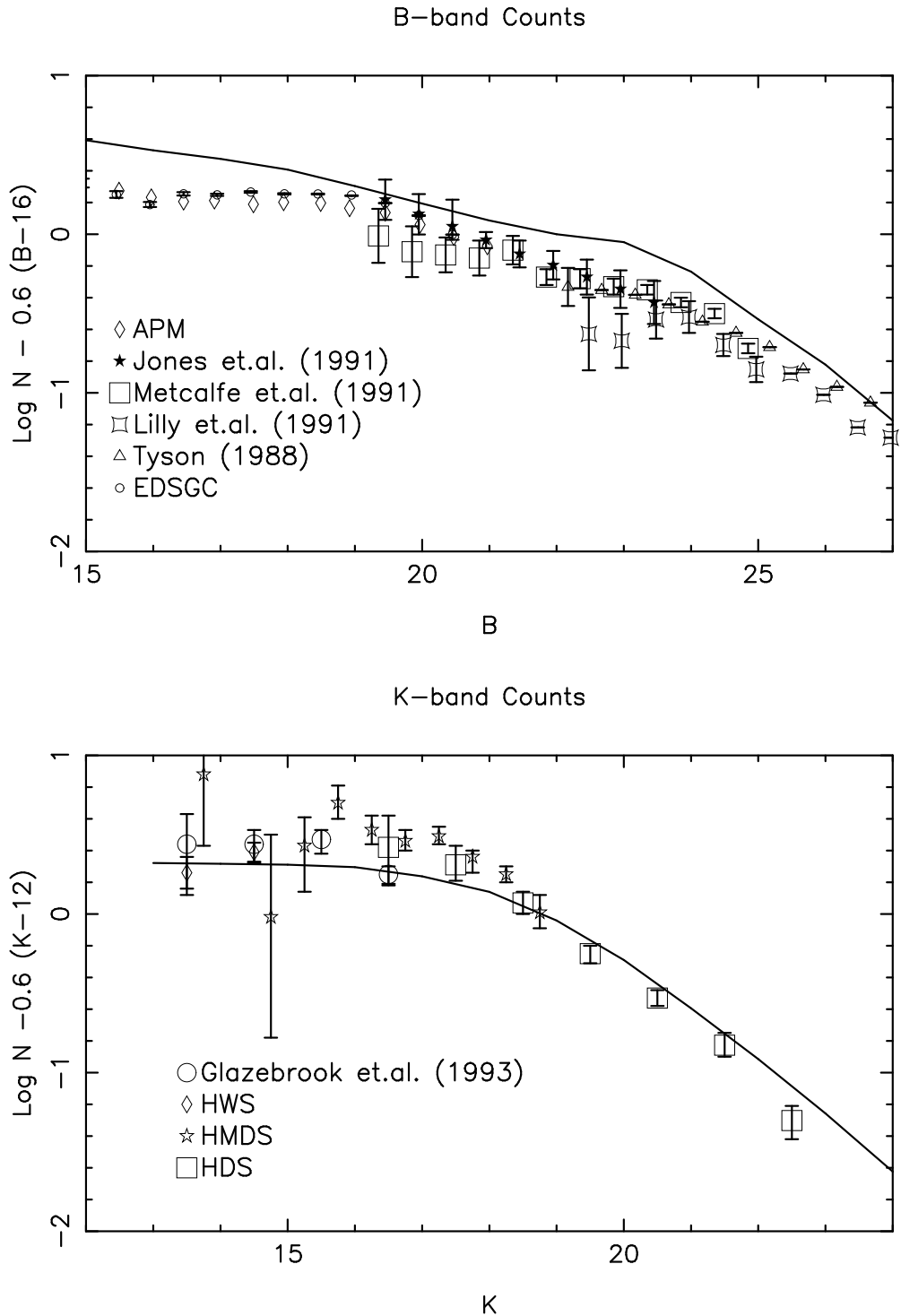

**Figure 18.** Differential galaxy number counts, N, per magnitude and per square degree. The data points are as in Figure 9. The solid line shows the prediction for a model which has the same parameters as the fiducial model except that the stars were assumed to form with the Miller-Scalo rather than the Scalo IMF (see Figure 4). The number counts are boosted above the prediction of the fiducial model since this choice of IMF causes galaxies at moderate redshifts to be brighter.

We next considered changing the overall shape of the IMF. Instead of using the Scalo IMF of the fiducial model, we switched to the Miller & Scalo (1979) function shown by the dashed curve in Figure 4. The H$\alpha$ luminosities are again affected since the two IMFs have different numbers of very high mass stars. However, the other properties of the present day galaxy population are, again, virtually unchanged. The robust nature of these results is very reassuring.

However, the $B$-band faint galaxy counts and associated redshift distribution are quite strongly affected by this change of IMF. Figure 17 shows that the redshift distribution is significantly deeper than that of the fiducial model. This behaviour is consistent with the passive evolution of the two different stellar populations illustrated in figure 5. Relative to the luminosity at late times ($t \gtrsim 10\mathrm{Gyr}$), the Miller-Scalo IMF produces more luminosity for the first $2-3\mathrm{Gyr}$. Together with the relatively late epoch of galaxy formation in these models, this implies that galaxies at moderate redshift are brighter with the Miller-Scalo IMF than they are in the fiducial model. This added "luminosity evolution" deepens the redshift distribution and also boosts the faint galaxy number counts as can be seen in Figure 18.

### 4.3 PREDICTIONS

In §4.1 we made a thorough comparison of a fiducial model with existing data on galaxy populations. We now present some predictions of this fiducial model which may be testable by observations in the not too distant future.

#### 4.3.1 Evolution of the Galaxy Luminosity Function

Our fiducial model predicts substantial evolution of the galaxy luminosity function at look-back times accessible with current observational techniques. In the observer's $B$- and $K$-bands there are four distinct processes that determine the evolution of the luminosity function: the $K$-correction, the passive evolution of stellar populations, star formation, and galaxy merging. The $K$-correction is most severe in the blue where at redshifts, $z \gtrsim 1$, the rest frame ultraviolet, produced predominately by young stars and stars in advanced stages of stellar evolution, is shifted into the $B$-band. The evolution observed in the $B$-band can therefore be quite different to that observed in the $K$-band.

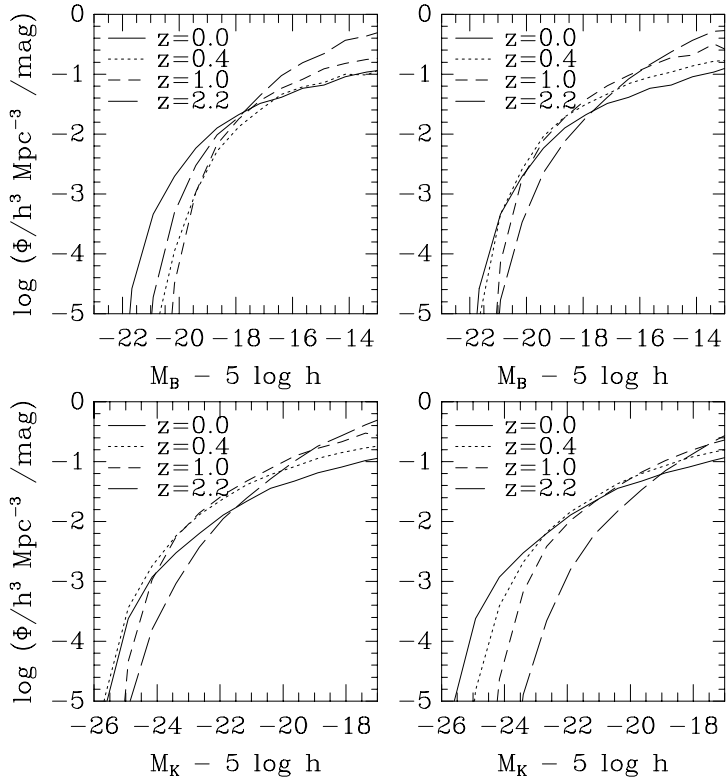

**Figure 19.** Galaxy luminosity functions in the observer's and galaxies' rest-frame $B$- and $K$-bands at various redshifts. The top row shows the observer's and rest-frame $B$-band luminosity functions respectively. The bottom row shows the corresponding $K$-band luminosity functions. The solid lines show the luminosity functions at $z = 0$ reproduced from Figure 6, while the dotted, short-dashed, and long-dashed lines show the predictions at $z = 0.4, 1.0$ and $2.2$ respectively.

Figure 19 shows our predicted $B$- and $K$-band luminosity functions at various redshifts in the observer's frame and in the rest-frame of the galaxies. The differences between them demonstrate the varying importance of the K-corrections. In the rest frame $K$-band, the characteristic luminosity, $L^\star$, fades gradually with increasing redshift, while the faint end slope of the luminosity function becomes progressively steeper. In the rest-frame $B$-band, the evolution is more complicated. Between $z = 0$ and $z = 0.4$, $L^\star$ remains approximately constant and the evolution is best characterized as a steady increase with lookback time of the number density of all galaxies fainter than $L^\star$. At higher redshifts, the shape of the luminosity function begins to evolve rapidly. By $z = 1$, the characteristic luminosity has decreased significantly and the faint end slope has become dramatically steeper. In the observer's frame the evolution is not even monotonic due to the changing influence of the K-corrections.

### 4.3.2 The Galaxy Redshift Distribution at Faint Magnitudes

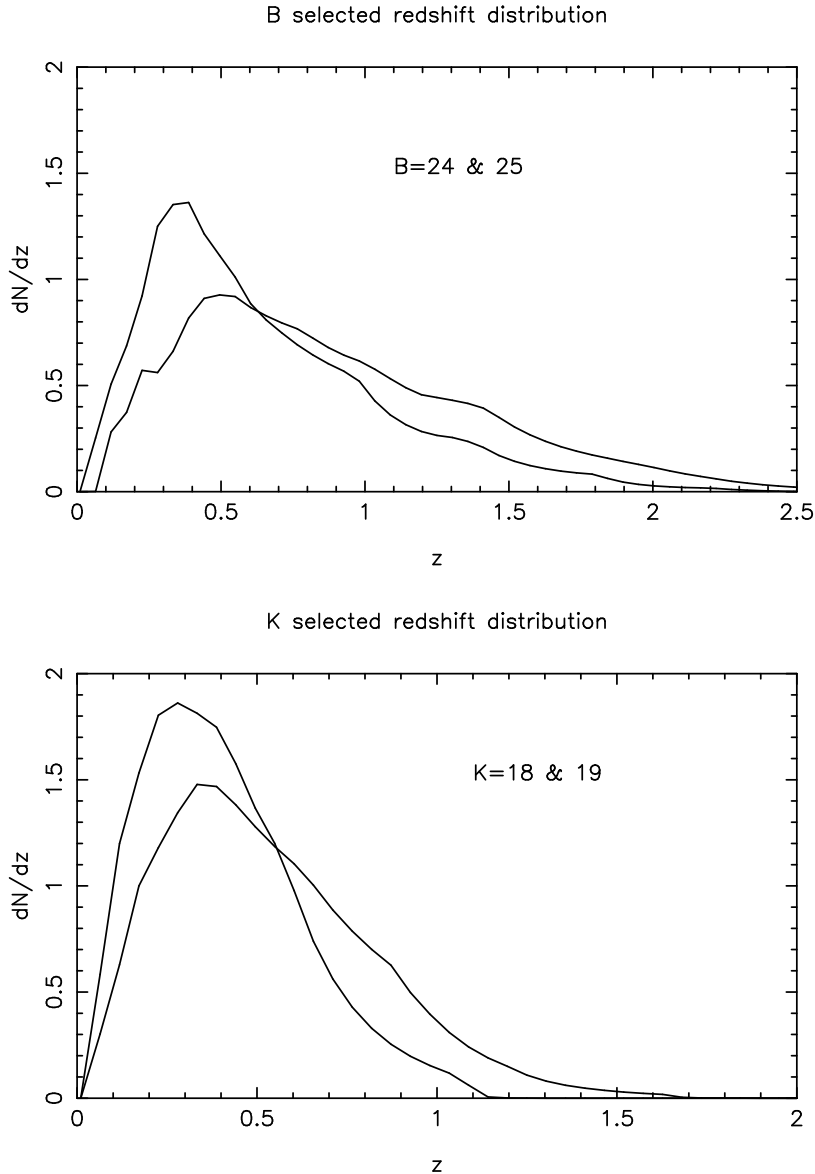

**Figure 20.**  The galaxy redshift distribution, $dN/dz$, predicted by the fiducial model of §4.1, for surveys limited at faint apparent magnitudes in the $B$- and $K$-bands.

The redshift distribution of faint galaxies places severe constraints on models of galaxy evolution. Our fiducial model is in excellent agreement with currently published data reaching to $B \approx 22$ (Colless *et al.* 1993). With the advent of multi-slit spectroscopy, it is now becoming possible to obtain near complete redshift catalogues at even fainter magnitudes. Surveys are currently underway at $B \approx 24$ (Glazebrook *et al.* in preparation) and $I \approx 22$ (Lilly *et al.* in preparation). We therefore present the distributions predicted by our fiducial model at faint apparent magnitudes. Figure 20 shows the redshift distributions of optically and infrared selected samples. In the infrared the evolving redshift distributions remain quite shallow even at $K = 18$ and 19. This is not the case in the $B$-band where a tail of high redshift galaxies is predicted. This tail appears when the UV flux produced by young stars in these star-forming galaxies is redshifted into the optical band.

### 4.3.3 The Star Formation History of the Universe

It may be possible in future to constrain the global star formation history in the universe using QSO absorption line systems as a probe of the formation of metals (see Lanzetta *et al.* 1993 and Pettini *et al.* 1993). In our fiducial model the total mass density that is converted into stars by the present, expressed in units of the critical density, is $\Omega_\star = 0.011$. In this model $\Omega_b = 0.06$ and so only 18% of baryons have been transformed into stars. Figure 21 shows how the mass in stars grows with time. Star formation begins at high redshift $z \gtrsim 4$, but the majority of stars are formed recently with half the stars being formed after $z \approx 0.9$. The current star formation rate averaged over all galaxies can be expressed as $\dot{\Omega}_\star/H_0 = 0.0013$; thus, at the present rate, a mass density corresponding to $\Omega_\star = 0.0013$ or $3.7 \times 10^8 h^2 M_\odot/\mathrm{Mpc}^3$ will be converted into stars in a Hubble time.

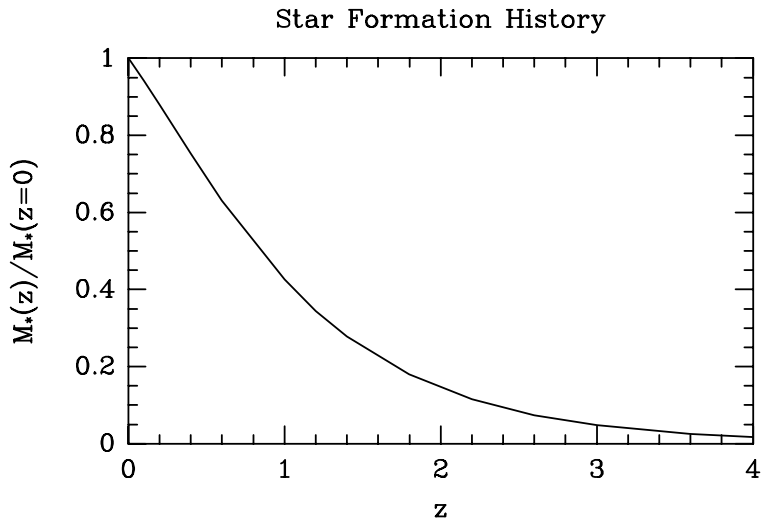

**Figure 21.** The history of star formation in the fiducial model of §4.1. The curve shows the mass fraction of stars formed by redshift $z$ relative to the present mass in stars.

### 4.3.4 Dark Baryons

At the present epoch, 82% of the baryons are not in stars. Most of these baryons (57% of the total) are associated with halos of mass $M_H < 10^{10} M_\odot$ which, in our model, eject all their gas after converting only a tiny proportion into stars. The remaining 25% of the baryons are in the form of cold, star forming gas and hot diffuse gas, either confined in galaxy clusters and galaxy halos, or simply expelled from lower mass field galaxies. Some of this gas will be hot enough to emit X-rays, but much of it is likely to be invisible. This gas and the mass in brown dwarfs implied by the value of $\Upsilon = 2.7$ constitute "baryonic dark matter" in our model.

## 5 Discussion and Conclusions

We have presented an *ab initio* model of galaxy formation in which we attempt to understand the formation paths and present day properties of galaxies starting from a spectrum of primordial density perturbations. Our efforts are hampered by the fact that several of the relevant physical processes are known only in a sketchy way. The approach we have adopted is to characterize these processes using scaling laws, inspired by simple physical considerations or by the results of numerical simulations.

Our recipe for galaxy formation is quite general and can readily be adapted to hierarchical schemes with any initial conditions. In this paper, we adopted the standard cold matter theory as the underlying cosmogonic model. Within this framework, we constructed a "fiducial" model by adopting reasonable values for the model parameters. These are two "feedback" parameters (the efficiency with which energy from supernovae and stellar winds is coupled to the intragalactic gas and the normalization of the star formation rate); two parameters that describe the stellar populations that form (the stellar initial mass function and the global stellar mass-to-light ratio); and two parameters that control galaxy mergers (a timescale and its dependence on mass). Having fixed the fiducial values of these parameters so as to obtain a reasonable match to the bright end of the $B$-band luminosity function, we calculated, and compared carefully with observations, predictions for a number of observables, viz, (i) the $K$-band luminosity function; (ii) $B-K$ colours as a function of absolute magnitude; (iii) H$\alpha$-line luminosity; (iv) the correlation between I-band magnitude and circular velocity; (v) the distribution of galaxy counts in the $B$- and $K$-bands as a function of apparent magnitude; (vi) the redshift distribution of the faint counts; (vii) the evolution of the luminosity function; and (viii) the history of galactic star formation. Some of these diagnostics are well determined observationally, but others, particularly (vi)-(viii), are genuine predictions of our model which may be tested with future observations. Finally, we explored the effect of varying the model parameters in order to assess the relative importance of the different physical processes that we have modelled.

Our models are strongly constrained by our combination of observed galaxy properties. Their main features may be summarized as follows:

**1)** The shape of the galaxy luminosity function is primarily controlled by the combined effect of galaxy mergers and stellar feedback. For plausible stellar mass-to-light ratios, the $B$- and $K$-band luminosity functions predicted at the present day are in reasonable agreement with observations. Although this represents a substantial improvement over the results of Cole (1991), White & Frenk (1991), and Kauffmann *et al.* (1993), our fiducial $B$-band luminosity function is still somewhat steep at absolute magnitudes fainter than about $M_B = -18$. The faint end slope could be flattened further, in closer agreement with the observed field luminosity function, by increasing the strength of the stellar feedback (as noted also by Lacey *et al.* 1993). However, we remain sceptical about the need to produce such a flat faint end slope. Deep surveys in Virgo (Sandage *et al.* 1985; Impey *et al.* 1988) and Fornax (Phillipps *et al.* 1987; Ferguson & Sandage 1988) reveal luminosity functions which are as steep or even steeper than that of our fiducial model, as do field studies of blue galaxies (Shanks *et al.* 1990). As Phillipps *et al.* have emphasized, field surveys could be biased against faint, low surface brightness, galaxies.

**2)** The colours and H$\alpha$ luminosities (a measure of the star formation rate) of our galaxies are most sensitive to our assumed baryon density and star formation timescale. Even neglecting metallicity effects, our colour-magnitude relation shows *no trend* for brighter galaxies to be bluer. Inclusion of metallicity effects would likely lead to our brighter galaxies being redder than our faint ones, as observed. This is a counter-intuitive result –in hierarchical models one naively expects more massive galaxies to be younger and bluer than fainter ones. In our model this trend is eliminated because massive galaxies tend to form through mergers of fragments which formed stars early and because their star forming activity is quenched when they lose their reservoirs of hot gas as they, preferentially, fall into larger potential wells. Our models therefore demonstrate that the colour-magnitude relation of elliptical galaxies is not a serious objection against the sequence of cosmogony in hierarchical models. This we regard as a major success of our model. Note, however, that quite specific conditions are required: in the models of Lacey *et al.* (1993) the brightest galaxies do tend to be the bluest.

**3)** Although our fiducial galaxies span most of the observed range of colours and H$\alpha$ luminosities, the reddest galaxies are not as red as many observed ellipticals by about 0.3 magnitudes in $B-K$. This is a serious shortcoming which cannot be remedied by reasonable changes in parameter values nor, it seems, by appealing to metallicity effects. Rather, it appears to be related to an age problem – the model does not produce galaxies with sufficiently old stellar populations. Indeed, about half the total number of stars are formed only since a redshift of 1. A modest increase in the available time after the beginning of structure formation would alleviate this problem, but this is not a parameter which we are at freedom to adjust within our cosmogonic framework. In a sense, this age problem is similar to the age problem for globular clusters in an $\Omega = 1$, $H_0 = 50\,\mathrm{km\,s^{-1}Mpc^{-1}}$ universe, not a surprising conclusion given that the same stellar evolutionary calculations underly both age determinations. In assessing the importance of this problem we must bear in mind that significant uncertainties remain in our understanding of the late stages of stellar evolution.

**4)** Our models give an excellent match to the observed number counts of faint galaxies as a function of $B$ and $K$ magnitude and the measured redshift distribution to $B \approx 22.0$. There has been much debate recently on the critical role that mergers may play in resolving the apparent contradiction between the large observed number of faint blue galaxies and their relatively shallow redshift distribution (Koo 1989; Rocca-Volmerange & Guiderdoni 1990; Guiderdoni & Rocca-Volmerange 1991; Broadhurst *et al.* 1992). In these studies, ad hoc merger rates were assumed and adjusted specifically to solve the "blue counts" problem. Galaxy mergers are, of course, innate to our models and have a dominant effect on the counts. Somewhat surprisingly, we find that the predicted $B$-band counts and associated redshift distributions are sensitive to the assumed shape of the IMF. (The $K$-band data are also affected by a change in the IMF, but much less so.) In contrast with earlier work (*e.g.* White & Frenk 1991, Lacey *et al.* 1993, Kauffmann *et al.* 1994), our models give an acceptable match to the faint counts even though the faint end slope of the present day luminosity function is fairly flat. If this slope were flattened even further by increasing the feedback strength, the number counts would drop slightly but this could be compensated for by a modest change in the IMF, to something intermediate between the Scalo and Miller-Scalo IMFs, with no appreciable effect on the remaining observables. The reason why our models can simultaneously give acceptable faint counts and a relatively shallow present-day luminosity function is simply that the luminosity function evolves substantially with time through the combined effect of mergers and feedback.

**5)** Our predicted $I$-band "Tully-Fisher" relation – the correlation between $I$-magnitude and circular velocity – has a slope and scatter in reasonable agreement with observation. However, our predicted zero-point for spiral galaxies is about 1.8 magnitudes too faint or, alternatively, our predicted circular velocities, at fixed magnitude, are about 60% too large. We regard this as the most serious shortcoming of our model, one which was previously noted by White *et al.* (1987), Cole (1991), White & Frenk (1991), Lacey *et al.* (1993), and Kauffmann *et al.* (1993) and which cannot be removed by adjusting model parameters. As Kauffmann *et al.* emphasize, perhaps the most instructive way to interpret this discrepancy is as an overabundance of dark galactic halos in the standard CDM cosmogony. Our model shows that the merger rate which gives an acceptable luminosity function and faint galaxy counts is insufficient to remove the excessive number of halos.

**6)** Since the properties of our model galaxies are known at all epochs, we can make definite predictions about the galaxy populations expected at intermediate and high redshift. Much of the activity associated with star formation and galaxy merging occurs at redshifts less than 1. In particular, we expect substantial evolution of the luminosity function even at quite modest redshifts, $z \lesssim 0.4$, but this evolution can be quite complex and depends sensitively on the band-pass in which the observations are carried out. We are also able to predict the redshift distributions expected in surveys at increasingly faint magnitudes. Only at $B \simeq 25$ do we expect a non-negligible tail of galaxies with $z \geq 2$, but even then, the predicted median redshift is relatively modest, $z \approx 0.7 - 0.8$. In the $K$-band it would be necessary to go to $K = 19$ to detect an appreciable population beyond redshift 1.

**7)** An important prediction of our model is the existence of "dark" baryons. Only about 20% of the baryons at the present epoch are locked up in stars, two thirds of which have mass less than $0.1\mathrm{M}_\odot$ (*ie* brown dwarfs, "Jupiters", etc.) The remainder are still in gaseous form, mostly in a hot intergalactic phase, but with a non-negligible fraction in a cool phase within star-forming galaxies. Thus, in principle, our models have no difficulty accounting for the large amount of cool gas inferred at $z \simeq 2$ from the study of damped Ly$\alpha$ clouds seen along the line-of-sight to many quasars (Wolfe 1988).

Many of the essential features of our model are also present in the Kauffmann *et al.* (1993) study, but the two implementations are very different. Thus, the Monte-Carlo methods, the feedback mechanisms, the star formation algorithms, and the galaxy merger rates are substantially different in the two approaches, as are the detailed properties we each chose to emphasize. Our cosmological model also assumes lower values of the baryon density and of the biasing parameter. On the whole, the results of the two studies are consistent with each other and the differences that we do find can readily be understood in terms of different parameter values and/or specific prescriptions for modelling key physical processes. For example, the flatter faint-end luminosity functions that we obtain result mainly from the stronger feedback and the higher merger rate of faint galaxies relative to bright ones in our model. Our lower baryon density and correspondingly longer cooling time for the gas is largely responsible for the differences at the bright end of the luminosity function and, together with our galaxy merger rates, explain why there is no trend for the brightest galaxies in our model to be the bluest. The main shortcoming of this kind of models is present in both studies although it is expressed in different ways. Kauffmann *et al.* fixed their parameter values so as to match the properties of the Milky Way and its satellites and thus, by implication, the zero-point of the Tully-Fisher relation. As a result the mean $B$-band luminosity density in their model turned out to be a factor of 2 too high. We, by contrast, fixed our parameters so as to match the $B$-band luminosity function near the characteristic luminosity, $L_*$, and thus obtained an acceptable luminosity density at the expense of an incorrect zero-point for the Tully-Fisher relation.

To summarize, we have laid down a sequence of steps leading to a detailed model of galaxy formation, starting from an assumed spectrum of primordial density fluctuations. Our model must be credited with a number of significant successes, most notably its ability to match, approximately, the present day galaxy luminosity function, the colour-magnitude relation, and the counts and redshift distribution of faint blue and red galaxies. This model, however, suffers from two major failings: its inability to produce bright galaxies as red as many observed ellipticals and to match the zero-point of the Tully-Fisher relation. These two problems can be traced back to the very core of the model – they cannot be removed simply by changing parameter values and will require a revision of some of the ideas developed here. Whether this revision will encompass only astrophysics (gas dynamics, star formation and stellar evolution) or whether it will necessitate a modification of the cosmological model (the values of $\Omega_0$, $\Lambda_0$, and $H_0$ or the input power spectrum) remains, at the present time, an open question.

**Acknowledgements**

We wish to thank Gustavo Bruzual and Stephan Charlot for providing us with their population synthesis model and associated software; Karl Glazebrook and Jon Gardner for supplying their compilations of faint galaxy number counts; Paul Young for allowing us to plot his Tully-Fisher data prior to publication; and John Lucey for supplying us with the data from his survey of Coma ellipticals. We are also grateful to Guinevere Kauffmann, Cedric Lacey and Simon White for their careful reading of the manuscript and their many valuable comments. This work was supported by grants from SERC.